\documentclass[twocolumn,letterpaper]{IEEEAerospaceCLS}  

\usepackage{cite}
\usepackage{caption}
\usepackage{amsmath,amssymb,amsfonts}
\usepackage{float}
\usepackage{subfig}
\usepackage{algorithm}
\usepackage{algpseudocode}
\usepackage{enumitem}
\usepackage{comment}
\usepackage[hidelinks]{hyperref}
\usepackage{soul}

\usepackage[]{graphicx}    
\PassOptionsToPackage{hidelinks}{hyperref}
\usepackage{hyperref}

\newcommand{\ignore}[1]{}  
\newcommand{\alg}[1]{\texttt{#1}}

\begin{document}
\title{Resilient UAV Data Mule via Adaptive Sensor Association under Timing Constraints}

\author{Md Sharif Hossen, Anil Gurses, Ozgur Ozdemir, Mihail Sichitiu, Ismail Guvenc \\ Department of Electrical and Computer Engineering, North Carolina State University, Raleigh, USA \\ Email: \{mhossen, agurses, oozdemi, mlsichit, iguvenc\}@ncsu.edu 
\thanks{\footnotesize 979-8-3315-7360-7/26/$\$31.00$ \copyright2026 IEEE}}

\maketitle

\thispagestyle{plain}
\pagestyle{plain}

\begin{abstract}
Unmanned aerial vehicles (UAVs) can be critical for time-sensitive data collection missions, yet existing research often relies on simulations that fail to capture real-world complexities. Many studies assume ideal wireless conditions or focus only on path planning, neglecting the challenge of making real-time decisions in dynamic environments. To bridge this gap, we address the problem of adaptive sensor selection for a data-gathering UAV, considering both the buffered data at each sensor and realistic propagation conditions. We introduce the \alg{Hover-based Greedy Adaptive Download (HGAD)} strategy, designed to maximize data transfer by intelligently hovering over sensors during periods of peak signal quality. We validate \alg{HGAD} using both a digital twin (DT) and a real-world (RW) testbed at the NSF-funded AERPAW platform. Our experiments show that \alg{HGAD} significantly improves download stability and successfully meets per-sensor data targets. When compared with the traditional \alg{Greedy} approach that simply follows the strongest signal, \alg{HGAD} is shown to outperform in the cumulative data download. This work demonstrates the importance of integrating signal-to-noise ratio (SNR)-aware and buffer-aware scheduling with DT and RW signal traces to design resilient UAV data-mule strategies for realistic deployments.
\end{abstract}
    
\tableofcontents

\section{Introduction}\label{sec:intro}
Aerial platforms that provide flexible, reliable, and autonomous communication support are becoming increasingly necessary as mobile networks extend into mission-critical areas, including tactical communications, emergency response, and battlefield intelligence. Unmanned aerial vehicles (UAVs) have become a viable option for wireless data collection from geographically dispersed sensors \cite{Liu2025}, as they offer better line-of-sight (LoS) links and provide good coverage with rapid deployment capabilities. In time-sensitive missions, UAVs can be used as a data mule (DM) to collect data from multiple sensors within strict mission durations and energy constraints \cite{zeng} \cite{Merwaday}.

In many real-world settings, the UAV trajectory is predetermined, but deciding which sensor to associate with at each time step is an open decision-making challenge. Each sensor is given a specific amount of data, and environmental obstacles, multipath fading, and UAV mobility can all cause considerable variations in wireless channel conditions. Given these limits, it is critical to design intelligent and real-time solutions that allow the UAV to maximize download efficiency while meeting data demands. Most prior work on UAV data collection has been on simple heuristic selection algorithms or path planning under static channel assumptions. For example, Zeng et al.~\cite{zeng_et_al} optimize UAV relaying for maximizing throughput, while Wu et al.~\cite{Wu} jointly optimize the multiuser communication scheduling and UAV trajectory over a finite horizon to maximize the throughput. Additionally, Liu et al.~\cite{Liu} optimize UAV trajectory jointly with scheduling under idealized channel models to minimize the flight time and maximize the data collection. These works, however, fall short in capturing the intricacies of actual wireless environments, where signal quality varies with distance and time. Furthermore, existing approaches overlook the need for stable sensor associations under favorable conditions and data buffer restrictions. Also, they have considered either offline trajectory planning or unrealistic assumptions regarding wireless channel behavior, often considering static environments or idealized signal conditions. Such strategies do not reflect the significant variability and dependency of signal dynamics experienced in actual field deployments.

To fill this gap, we propose an adaptive, buffer-aware sensor association for UAV-based data gathering based on realistic signal traces collected from the NSF AERPAW digital twin (DT) and real-world (RW) testbed. This enables us to simulate UAV decision-making based on environments that mimic actual wireless propagation conditions, including terrain-aware signal-to-noise ratio (SNR) fluctuations.

Our main contribution is the \alg{Hover-based Greedy Adaptive Download (HGAD)} approach, which enables the UAV to adaptively hover close to a sensor upon observing strong SNR and favorable throughput conditions and maximizing the throughput within mission time constraints. This is opposed to a conventional \alg{Greedy} heuristic approach, which switches sensors based on instantaneous SNR values and is susceptible to instability, high handover rates, and an inefficient buffer. With the addition of buffer awareness, SNR awareness, and hover-based logic, \alg{HGAD} facilitates more stable and effective buffer-aware downloads in both fixed and autonomous trajectories of UAV mobility.

We compare \alg{HGAD} with a baseline \alg{Greedy} solution through four operating modes: (i) a fixed path that mimics actual UAV flight traces in the DT environment, (ii) a fixed path mimicking the UAV flight in simulation, (iii) an autonomous path in simulation where the UAV self-adjusts based on the buffer states at each sensor, and (iv) a fixed path with two different flights in the RW testbed, which includes hardware, software, and a real-time environment. Our findings demonstrate that \alg{HGAD} not only enhances download stability and cumulative throughput but also achieves a more fair and effective use of all sensors, rendering it a realistic solution for mission-oriented UAV deployments. The contributions of this work are as follows:
\begin{itemize}[left=0em]\setlength\itemsep{0.5em}
\item We propose herein an adaptive approach, \alg{HGAD}, that enables the UAV to hover close to a sensor during the maximum throughput period, improving link stability and efficiency.
\item \alg{HGAD} jointly considers the instantaneous SNR and the data buffer states at each sensor simultaneously, which are usually overlooked in existing UAV data-gathering models.
\item We leverage actual signal traces from the NSF AERPAW DT and RW testbed, with realistic terrain-aware SNR variation. 
\item We compare fixed and self-navigating UAV flight paths to evaluate the performance of \alg{HGAD}. 
\end{itemize}

The remaining sections of the paper are arranged as follows. We discuss the literature review in Section \ref{sec:rw}.  In Section \ref{sec:sm}, the system model, along with the problem definition, is illustrated. The sensor selection techniques, i.e., \alg{Greedy} baseline and the \alg{HGAD}, along with the fixed and autonomous trajectories for the UAV DM, are described in Sections \ref{sec:ft_dm} and \ref{sec:at_dm}, respectively.
 In Section \ref{sec:data_collection}, we discuss the experimental setup and how we collect UAV trajectory telemetry data. Comparative results are explained in Section~\ref{sec:nr}, which compares the performance of \alg{Greedy} and \alg{HGAD} approaches. Section \ref{sec:con} includes the conclusion and future works.\label{intr}

\section{Related Work}\label{sec:rw}
UAVs have been studied extensively for data collection in wireless networks with mission-constrained and delay-sensitive applications. However, a significant portion of prior work focused on offline trajectory planning, considering a preplanned trajectory for the UAV flight to minimize distance, energy, or mission time. The drawback of these works is that there was no testbed or DT data to verify the performance in the real world. For example, in \cite{Liu}, the authors provide a simulation-centric UAV trajectory planning design without validating realistic signal data. The design also excludes adaptive SNR-based sensor association and data buffer constraints, and hence may be of limited use in a dynamic setting. The work targets a single optimized path with link assumptions and provides minimal insight into the real-time environments. 

Additionally, in \cite{Binol}, the authors apply evolutionary algorithms for multi-UAV path planning to gather data from roadside units (RSU) to minimize overall mission time. Their study considers various simulation abstractions and excludes real-world data; therefore, the performance in real-world deployments might be overestimated. Furthermore, their work overlooks the data buffer states of each sensor, assuming data exchange within the RSU's coverage, and fails to account for fluctuating link quality or SNR. Our approach combines buffer-aware adaptive sensor selection with realistic DT and RW signal traces, which provides a more practical signal-driven evaluation. Even though Krishnan et al.'s work \cite{Krishnan} uses traveling salesman problem (TSP)-based boundary optimization to reduce UAV flight distance in a mathematically elegant manner, it only works in a highly idealized simulated environment. The study does not address genuine signal fluctuation, connection degradation, or per-sensor data restrictions; instead, it assumes stable, circular communication zones and constant SNR. Furthermore, it ignores adaptive UAV tactics that are crucial in real-world deployments with DTs or terrain-aware environments, such as hovering or dynamic sensor selection.

 Besides UAV trajectory optimization, classic DM work has also been influential in formulating wireless data collection strategies. Sugihara and Gupta~\cite{Sugihara_Gupta} formulated the mobile DM path selection problem as a label-covering tour and designed approximation algorithms for minimum data delivery latency. Their results, verified in Matlab and ns2 simulations, estimated that controlled mobility would efficiently exploit communication ranges. These works, however, were based on idealized wireless models and offline planning and did not consider realistic signal variations. By comparison, our work supports the use of RW testbeds and DT traces of the physical world to design an adaptive sensor association approach that considers instantaneous SNR and buffer availability. It boosts the efficiency and robustness of UAV-enabled data aggregation.
Our work addresses this gap by introducing the \alg{HGAD} policy, where a UAV adaptively hovers close to a sensor in conditions of potential high throughput. We focus on digitally emulated wireless channel traces from the NSF AERPAW DT and RW testbeds, including the simulated UAV flight traces, to investigate the performance of \alg{HGAD} in various conditions. This hybrid evaluation bridges the gap between practical deployments and design principles, demonstrating a suitable and deployable UAV communications strategy.
\label{rw}

\section{System Model and Problem Formulation}\label{sec:sm}
We consider a general UAV-aided wireless data gathering system, where an aerial platform is employed as a \textit{data mule} to gather buffered data from geographically distributed multiple ground sensors or infrastructure nodes. The UAV operates in a constrained area (because of energy, regulatory, or mission constraints) and possesses a finite flight duration. The route can be precomputed (fixed path) or autonomously adapted in real time based on the link quality of the sensors. Each sensor is assigned an initial data buffer, and the data must be offloaded by the mission's end time. The UAV can establish a wireless connection to at most one sensor within its communication radius at every time instance. The achievable data rate depends on the instantaneous signal-to-noise ratio (SNR), which is a function of the UAV position, sensor distance, and propagation environment. Terrain, multipath fading, and UAV mobility are some of the environmental dynamics that cause a time-varying variation of this link quality.

\begin{figure}[t]
    \centering        \includegraphics[width=3.5in,trim=.2in 1.2in 0in 1.2in,clip]
    {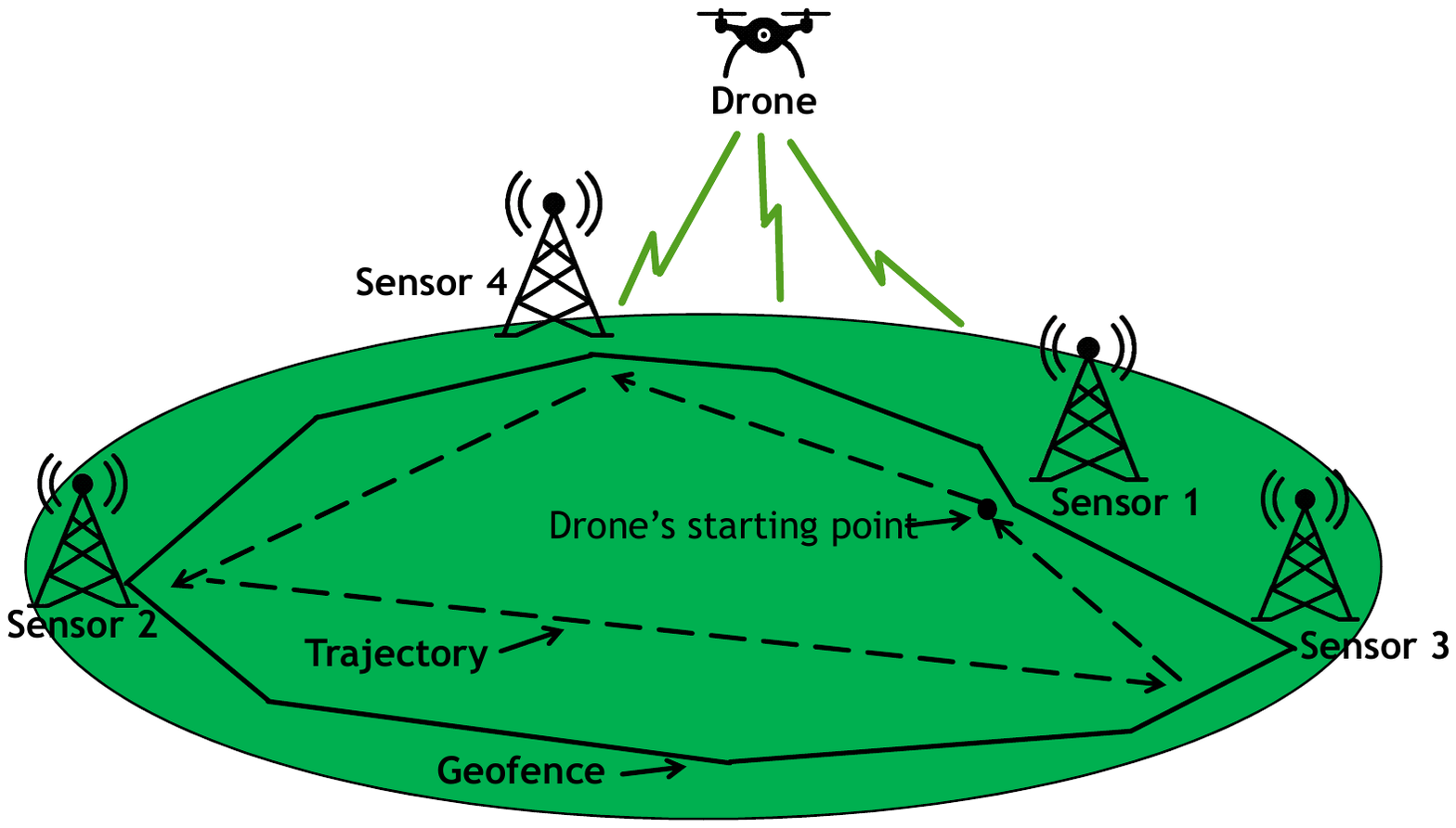}    
    \caption{Representative UAV flight trajectory and sensor positions.}
    \label{fig:sys_traj}    
\end{figure}
Fig.~\ref{fig:sys_traj} shows the system configuration. The dotted black line is the UAV flight path, which can be fixed or dynamically altered in real-time. The ground sensors each have data to buffer for upload. The geofence boundary illustrates regulatory or mission-imposed limits on the UAV mobility. As the UAV traverses its trajectory, it opportunistically establishes wireless links with sensors within communication range, with instantaneous SNR determining the effective throughput. Fig.~\ref{fig:sys_traj} also illustrates the interaction of spatial topology, sensor distribution, and constraints on UAV mobility that together constitute the adaptive sensor selection problem.

Let us consider a set of ground sensors $\mathcal{S}=\{1,\dots,\ N\}$ as shown in Fig.~\ref{fig:sys_traj}. Initially, the sensor $i$ is located at $s_i\in\mathbb{R}^2$, and has $Q_i>0$ bits of data in its buffer to be downloaded by the UAV. Time is divided into slots $t=0,1,\dots, T-1$ with a slot length $\Delta t>0$~seconds. The
location of the UAV at the slot $t$ is $r_t\in\mathbb{R}^2$. The UAV must be within a mission geofence $\mathcal{G}\subset\mathbb{R}^2$, follow a maximum speed $v_{\max}$, and start/end at locations $r_{\mathrm{start}}$ and $r_{\mathrm{end}}$. At most, a single sensor can be actively serviced in a slot (single-sensor connectivity). The instantaneous  $\mathrm{SNR}_i(r)$ for the UAV when it is at position $r$ and receiving data from the sensor $i$ is calculated as~\cite{Hossen}: 
\begin{equation}
\mathrm{SNR}_i^{\mathrm{dB}}(r)
= P_{\mathrm{tx}}^{\mathrm{dBm}}
+ G_{\mathrm{tx}}^{\mathrm{dBi}}
- \mathrm{PL}^{\mathrm{dB}}(r)
+ G_{\mathrm{rx}}^{\mathrm{dBi}}
- N_0^{\mathrm{dBm}},
\label{snr}
\end{equation}
where $P_{\mathrm {tx}}$ is the transmit power of a sensor, $G_{\mathrm {tx}}$ and $G_{\mathrm {rx}}$ are the antenna gains of the sensor and the UAV, respectively. $N_\mathrm{0}$ is the noise power, and $\mathrm{PL(\cdot)}$ is the distance-dependent path loss. 
SNR is mapped to data rate via a tabulated function $f(\cdot)$. The \emph{position-dependent per-slot capacity} (bits/s) is given as:
\begin{equation}\label{Eq:lookup_Table}
R_{i}(r) = f(\mathrm{SNR}_{i}(r)),
\end{equation}
\noindent
At each slot $t=0,1,\ldots,T-1$, we consider the \emph{time-varying} UAV position $r_t\in\mathbb{R}^2$, the binary association variables $x_{i,t}\in\{0,1\}$ (equal to 1 if the sensor $i$ is served in slot $t$), and the downloaded bits $y_{i,t}\ge 0$. The instantaneous per-slot capacity is evaluated at the \emph{current} UAV position, i.e., $R_i(r_t)$. The following constraints and state update govern the system:
\begin{align}
\sum_{i\in\mathcal{S}} x_{i,t} &\;\le\; 1,\quad \forall t,
\label{eq:single_assoc}\\
0 &\;\le\; y_{i,t} \;\le\; R_i(r_t)\,x_{i,t},\qquad \forall i,\,t,
\label{eq:capacity}\\
Q_{i}(t{+}1)&\;=\;\max\!\big\{\,Q_{i}(t)-y_{i,t},\,0\,\big\}, 
\label{eq:buffer}\\
Q_{i}(0)&=Q_{i}, \\
\lVert r_{t+1}-r_{t}\rVert &\le v_{\max}\Delta t, \quad t=0,\dots,T-2,\label{eq:mobility0}\\
r_0 &= r_{\mathrm{start}}, \quad r_{T-1}=r_{\mathrm{end}}, \quad r_{t}\in\mathcal{G},
\label{eq:mobility}
\end{align}
where $Q_i$ denotes the initial buffered data, and $Q_i(t)$ is the remaining data with $Q_i(0)=Q_{i}$.
The constraint in~\eqref{eq:single_assoc} enforces that the UAV can be connected only to a single sensor at a given slot; \eqref{eq:capacity} limits the downloaded bits by the distance-dependent capacity at the \emph{current} position $r_t$; \eqref{eq:buffer} updates each sensor’s remaining buffer without underflow; and \eqref{eq:mobility0}-\eqref{eq:mobility} restricts UAV motion (speed), start/end locations, and geofence coverage as $r_t$ changes over time.

Given a mission duration $T$, the maximum throughput is obtained as:
\begin{align}
\max_{\{r_t,x_{i,t}\}} \quad
& \sum_{i\in\mathcal{S}}\sum_{t=0}^{T-1} y_{i,t}, \label{Eq:maximize}\\
\text{s.t.}\quad\tag*{}
& \text{(\ref{eq:single_assoc}–\ref{eq:mobility})},\\
& \sum_{t=0}^{T-1} y_{i,t}\le Q_i\ \forall i, \label{Eq:RemainingData}
\end{align}
where the UAV chooses the waypoints $r_t$ and sensor associations $x_{i,t}$ to maximize the total throughput over the mission period. 

This formulation provides the foundation for adaptive sensor selection strategies, allowing the UAV to dynamically prioritize communication opportunities based on instantaneous SNR measurements and remaining data requirements (captured by~\eqref{Eq:RemainingData}). It ultimately optimizes both throughput efficiency and mission completion reliability. In the next two sections, we will study fixed and adaptive selection of $r_t$ for solving~\eqref{Eq:maximize}.

\section{Data Mule Operation under Fixed UAV Trajectory}\label{sec:ft_dm}
\begin{figure*}[!t]
    \centering
    
    \subfloat[]{\includegraphics[width=0.45\textwidth]
    {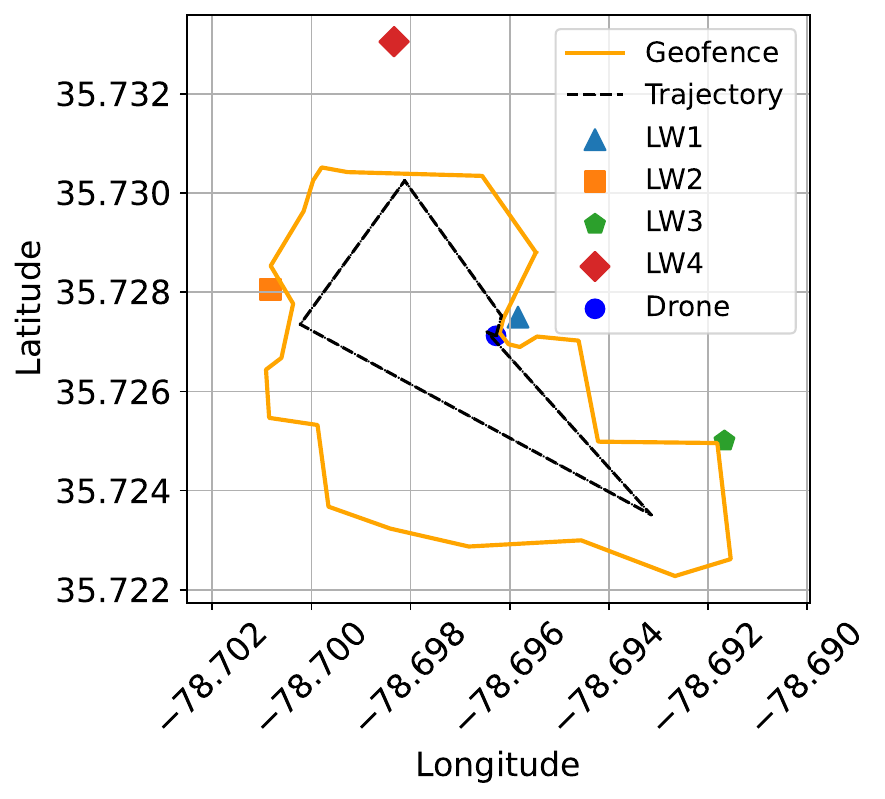}} 
    \subfloat[]{\includegraphics[width=0.45\textwidth]
{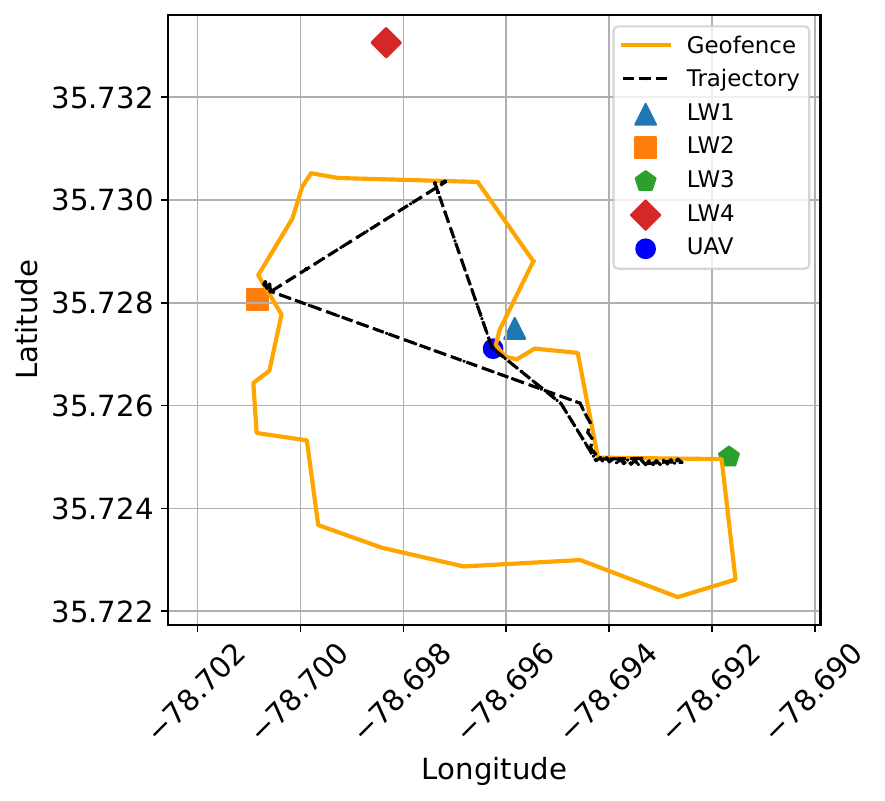}} 
    
    \caption{UAV as a data mule flight trajectory: (a) fixed trajectory from DT and (b) autonomous trajectory from simulation.}
    
    \label{fig:trajectory}
\end{figure*}

For the fixed trajectory case, as discussed in our previous work \cite{Hossen}, the UAV follows a pre-computed flight path before the mission begins. The UAV is not allowed to deviate from the trajectory during the mission. This case applies to situations where airspace constraints, energy constraints, or mission-dependent policies prevent dynamic trajectory deviation. In the fixed trajectory, we investigate two sensor selection methods: a baseline \alg{Greedy} approach and a proposed \alg{Hover-based Greedy Adaptive Download (HGAD)} strategy. Fig.~\ref{fig:trajectory}(a) shows the UAV flight mission for an example fixed trajectory. The idea is that the UAV will follow a predetermined path within the yellow-marked restricted area, also called a geofence, and download data from each sensor.

\subsection{Baseline Greedy Sensor Selection}
Using the baseline \alg{Greedy} approach, the UAV selects the sensor with the highest instantaneous SNR at each time step. At the beginning of the process, each sensor $i \in \mathcal{{S}}$ is assigned an initial data buffer $Q_i(0)$, and the cumulative data downloaded from each sensor is initialized as $D_{\mathrm{i}}(0) = 0$. 
The set of sensors $\mathcal{ C}(t) \subseteq \mathcal{S}$, which contains all sensors that UAV has finished downloading their buffered data, is initially empty, i.e., $\mathcal{C}(0) = \emptyset$.
At each time step, the UAV identifies the set of sensors with data in their buffers as $\mathcal{S}'(t) = \mathcal{S} \setminus \mathcal{C}(t)$. From this set, it selects the sensor $i^*(t)$, an index (or ID) of the strongest sensor at time step $t$,  that has the highest SNR, i.e.:
\begin{equation}
i^*(t) = \arg\max_{\mathrm{i} \in \mathcal{\mathrm S}'(t)} \text{SNR}_i(t).
\label{highest_snr}
\end{equation}
Throughput is then obtained from \eqref{Eq:lookup_Table}. The cumulative download and remaining data in sensor $i$'s buffer are then updated based on the data downloaded during slot $t$.  
If all the buffered data is downloaded from a sensor, i.e., 
$Q_{i}(t+1) \leq 0$, the sensor is marked as completed.
The time step counter is then incremented: $t \gets t + 1$, and the process continues until data from all sensors are downloaded. 
This strategy ensures that the UAV prioritizes the most favorable communication link at each step while respecting sensor-specific data buffers. 

\subsection{Hover-based Greedy Adaptive Download (HGAD) Strategy}
\alg{HGAD} increases data download efficiency by adapting the speed of the UAV  and the hovering approach based on the achievable data rate from each sensor. The core idea is to prioritize the strongest link at each time step while \textit{hovering near the sensor} when the maximum possible spectral efficiency is achieved. 
At each slot $t$, \alg{HGAD} selects the sensor with the largest instantaneous
achievable rate $R_i(r_t)$ defined in~\eqref{Eq:lookup_Table}. When the
UAV is near the best location of a sensor $i$, i.e.,~\eqref{Eq:lookup_Table} is close to its local maximum, the UAV hovers there to download data faster. The formulations are given as:
\begin{equation}
i^*(t) = \arg\max_{i \in \mathcal{S}(t)} R_i(t),
\label{Eq:sensor-selection}
\end{equation}
\begin{equation}
R_{i^\star}(r_t) \;\ge\; \gamma_i,
\label{eq:hgad-trigger}
\end{equation}
\begin{equation}
T^{\mathrm{hover}}_i = \min \left( \frac{D^{\mathrm{rem}}_i}{R^{\max}_i}, \; T^{\max}_i \right),
\label{eq:hovering-time}
\end{equation}

where $\gamma_i$ represents a maximum achievable data rate for a sensor $i$, which is obtained from the historical data. $T^{\mathrm hover}_i$ is the estimated hovering time for a sensor $i$, $D^{\mathrm rem}_i$ is the remaining data to download (in Mbits), $R^{\mathrm max}_i$ is the maximum data rate (in Mbits/sec) of a sensor, and $T^{\mathrm max}_i$ is the maximum time a UAV can hover near a sensor. If not hovering, the UAV iteratively selects the best sensor based on the sensor with the highest signal strength. 

Let us further clarify~\eqref{Eq:sensor-selection} -- \eqref{eq:hovering-time}. At every instant $t$, the UAV chooses the sensor $i$ with the highest instantaneous achievable data rate from the candidate pool, ensuring that the UAV does not waste time on the mission when operating on persistently weak links. Once hovering is initiated, the hovering time is set as the minimum time required to download the remaining data at the maximum achievable rate and within a predetermined bound. This assures that a lower rate cannot monopolize the mission time. Rather, the UAV focuses on maximizing the data download from the sensors with equal mission durations. For example, a UAV might get stuck downloading data from LW4, considering LW4's maximum link quality. $T^{\mathrm{hover}}_i$ will solve this issue by forcing the UAV to change its trajectory to download data from other sensors, which helps to maximize the total data download within the timing constraints.

\section{Data Mule Operation under Autonomous UAV Trajectory}\label{sec:at_dm}
Unlike the fixed trajectory case, for an autonomous trajectory, the UAV path is not pre-computed; instead, it adapts online to sensor status (whether a sensor has data to offload), link quality, and buffer size, optimizing motion and data transfer jointly~\cite{Hossen}. We implement two policies: a \alg{Greedy} SNR-based navigation and \alg{HGAD}.

\subsection{Baseline Greedy Sensor Selection}
Following the autonomous trajectory discussed in our previous work \cite{Hossen}, the baseline \alg{Greedy} approach for sensor selection is shown in Algorithm~\ref{alg:greedy_autonomous_trajectory}. The UAV at each decision point considers only sensors that have data remaining in their buffers. If it determines that all data has been downloaded from all sensors, it returns home. Otherwise, it selects the sensor with the highest instantaneous SNR among the available ones.
The loop continues until all sensor buffers are empty or the mission time ends. 

\begin{algorithm}[!t]
\caption{Baseline Greedy Sensor Selection}
\label{alg:greedy_autonomous_trajectory}
\begin{algorithmic}[1]
\State \textbf{Input:} Sensors, data buffer, SNR logs, UAV position trace
\State \textbf{Output:} UAV trajectory with download logs


\ForAll{time step $t$}
    \State Selects available sensors that have data to send

    \If{no data left to download from all the sensors}
        \State UAV returns to the home location
        \State \textbf{break}
    \EndIf

    \State Select a sensor $i$ with the highest SNR among candidates
    
    \If{total data download from a sensor $i$ $<$ initial data buffer of $i$}
        \State download data from the sensor $i$ current rate
        \If{no data left to download from sensor $i$}
            \State Sensor $i$ is marked as complete
        \EndIf
    \EndIf
    \State UAV moves towards the next waypoint
\EndFor
\end{algorithmic}

\end{algorithm}

\subsection{Hover-based Greedy Adaptive Download (HGAD) Strategy}
Algorithm~\ref{alg:hover_autonomos_trajectory} defines an autonomous UAV policy that considers both signal quality and the amount of data in each sensor's buffer, in such a way that the UAV maximizes the data collection from the ground sensors. At any time step, the UAV first checks whether all the sensors have already transferred their assigned data. If that is satisfied, the UAV returns to its home location. Otherwise, the UAV identifies the candidate sensors with remaining data in their buffers and selects the one with the largest remaining data. UAV then hovers near the chosen sensor if the SNR is high and downloads as much data as possible before the hovering duration ends, which is calculated from the throughput of the sensor at a given time slot. It then proceeds to the sensor with the next-largest remaining data to be downloaded in its buffer. If the UAV is in motion instead of hovering, it opportunistically downloads from the sensor among the candidates with non-empty buffers with the strongest SNR. If the current data rate drops below the maximum level possible in~\eqref{Eq:lookup_Table}, the UAV increases its speed towards a sensor (which has been discussed in our previous work in~\cite{Hossen} and the source code is available in~\cite{uavsimframework}) having maximum SNR and hovers there to download data. Throughout the procedure, the UAV records its position, download speed, and sensor status, maintaining an accurate trajectory log. This approach enables buffer-aware, SNR-based UAV navigation and supports an efficient mission planning for wireless data-gathering applications.

\begin{algorithm}[!t]
\caption{Hover-Aware Sensor Selection}
\label{alg:hover_autonomos_trajectory}
\begin{algorithmic}[1]
\State \textbf{Input:} Sensor, data buffer, SNR logs, UAV position trace
\State \textbf{Output:} UAV trajectory with download logs


\ForAll{time step $t$}    
    \State Selects available sensors that have data to send.
    \If{high data rate is not achieved at a sensor $i$}
        UAV flies at maximum speed
    \EndIf

    \If{no data left to download from all the sensors}
        \State UAV returns to its home location        
    \EndIf

    \State Select a sensor $i$ with the highest SNR among candidates
    
    \If{$hovering \gets \text{True}$ and total data download from a sensor $i$ $<$ initial data buffer of $i$}
        UAV downloads data at a high rate

        \If{no data left to download from a sensor $i$} sensor $i$ is marked as download complete and $hovering \gets \text{False}$
        \EndIf
    \Else 
        \For{the available sensors that have data to send, are chosen based on the SNR values}
            \If{total data download from a sensor $i$ $<$ initial data buffer of $i$}
                \State download data from sensor $i$ 
                \If{no data left to download from sensor $i$} sensor $i$ will be marked as download complete
                \EndIf
            \EndIf
            \If{a sensor $i$ has higher data buffer} $hovering \gets           \text{True}$ and UAV hovers near sensor $i$
            \EndIf                
        \State \textbf{break}            
        \EndFor
    \EndIf
    \State UAV moves towards the next waypoint with maximum speed
\EndFor
\end{algorithmic}
\end{algorithm}

\section{Experimental Setup and Wireless Dataset Collection}\label{sec:data_collection}
We employ three different modes of experiments, namely simulation, DT, and the RW testbed, to evaluate the performance of the autonomous data mule approaches introduced in the earlier sections. 

We first apply the \alg{HGAD} algorithm in a Python-based simulation environment. The controlled environment is used to quickly prototype and test the algorithm under ideal channel conditions with a free-space path loss model. Second, we evaluate \alg{HGAD} in a DT environment. The DT employs emulated signal traces and realistic radio channel models while executing the exact software that will eventually get deployed in the testbed.
Finally, we evaluate \alg{HGAD} on the AERPAW RW testbed, which consists of a multi-rotor UAV in combination with an AERPAW portable node carrying a USRP B205mini software-defined radio (SDR), while the ground BSs are equipped with USRP B210 SDRs.
The UAV establishes wireless links with ground SDR nodes while following predetermined trajectories and commanded altitudes. Operating in sub-6 GHz bands, the SDRs enable measurements of interference effects, throughput dynamics, and SNR variations. Unlike simulations or digital twins, real-world experiments capture nonidealities such as UAV flight dynamics, hardware constraints, and environmental factors like multipath and shadowing. Within this deployment, we validate \alg{HGAD} under realistic conditions and demonstrate its robustness and adaptability to inherent uncertainties in the field.

\begin{figure*}[t]
    \centering

    \subfloat[]{\includegraphics[width=1.8in]{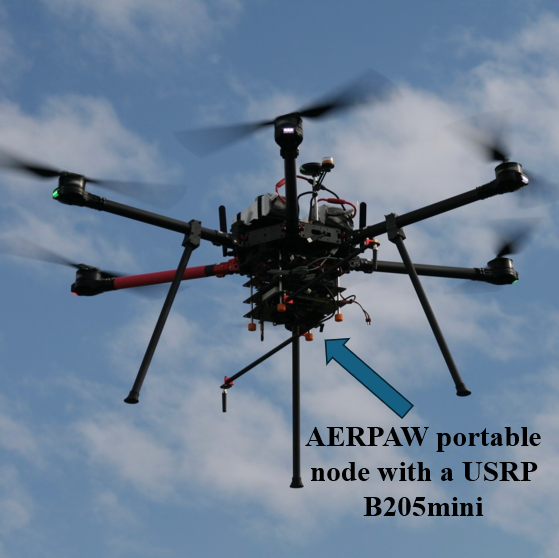}}
    \hspace{0.2in}
    \subfloat[]{\includegraphics[width=2.1in]{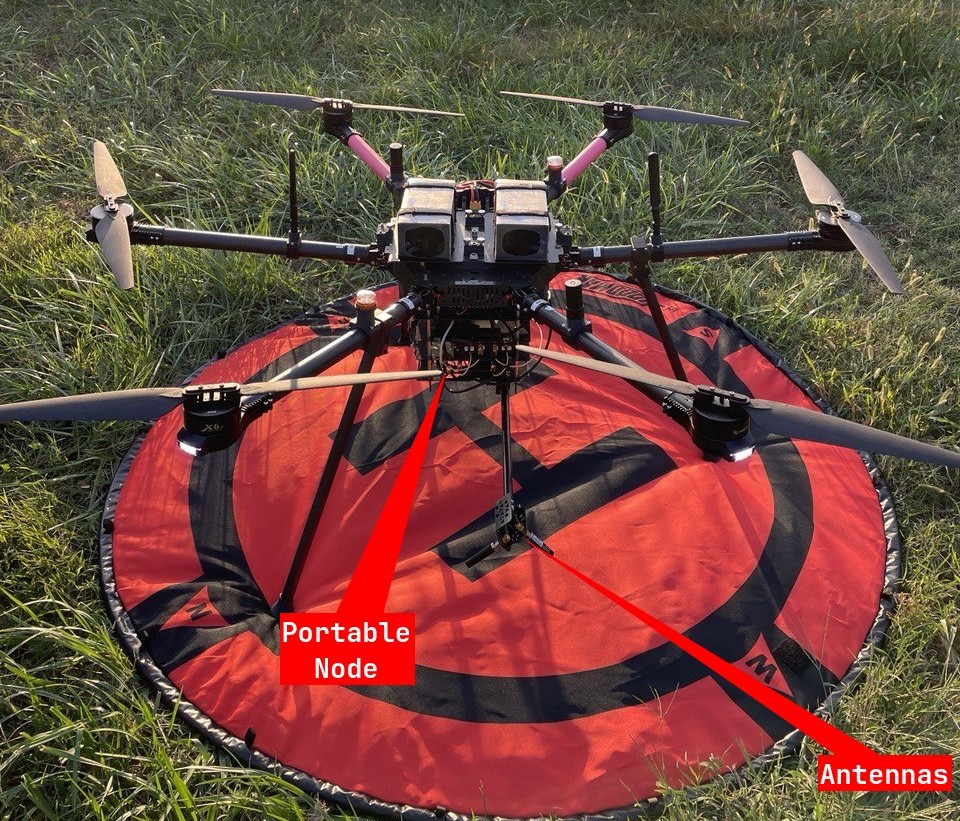}} 
    \hspace{0.2in}
    \subfloat[]{\includegraphics[width=2.14in]{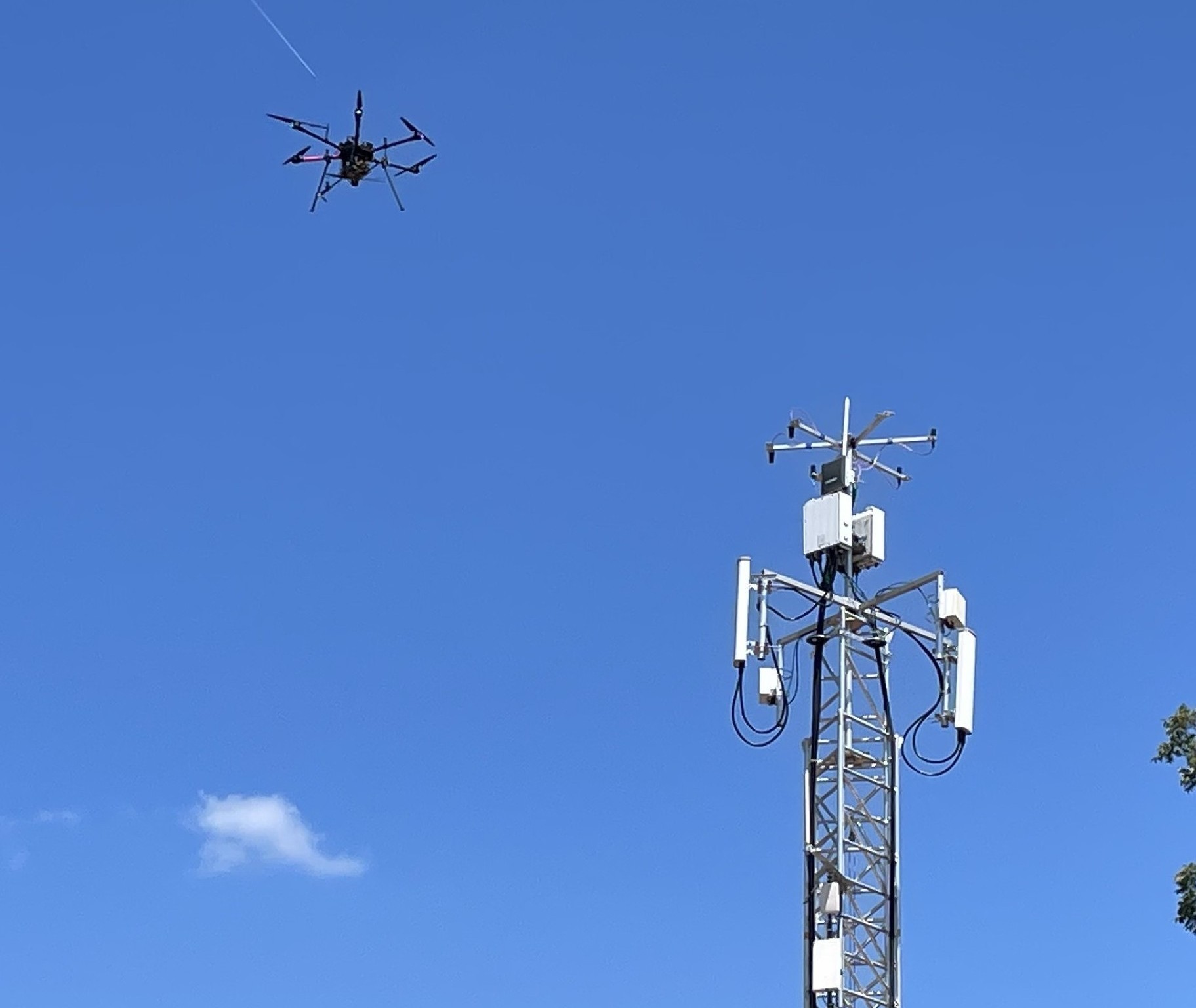}} 
    

    \caption{Experiment setup: (a) AERPAW UAV with SDR portable node, (b) UAV with USRP B205mini equipped portable node on the ground before flying, (c) UAV flying near a BS in the AERPAW testbed.}
    \label{fig:testbed_setup}
\end{figure*}

Fig.~\ref{fig:testbed_setup} shows the experimental setup used in the AERPAW testbed. Fig.~\ref{fig:testbed_setup}(a) shows the UAV flying in the sky, holding the portable node with a USRP B205mini mounted underneath. Fig.~\ref{fig:testbed_setup}(b) shows the UAV in closer proximity to the ground in preparation for takeoff, with the portable node and SDR hardware visible on the platform to give an unobstructed view of the instruments used to capture and transmit signals. Fig.~\ref{fig:testbed_setup}(c) shows the UAV in flight in the proximity of an AERPAW base station (BS), which serves as a sensor to download data from. 
These figures provide a broad overview of the RW experiment, from UAV-SDR integration to the practical RW deployment in the AERPAW testbed.

\begin{figure*}[t]
    \centering
    \subfloat[]{\includegraphics[width=2.7in]{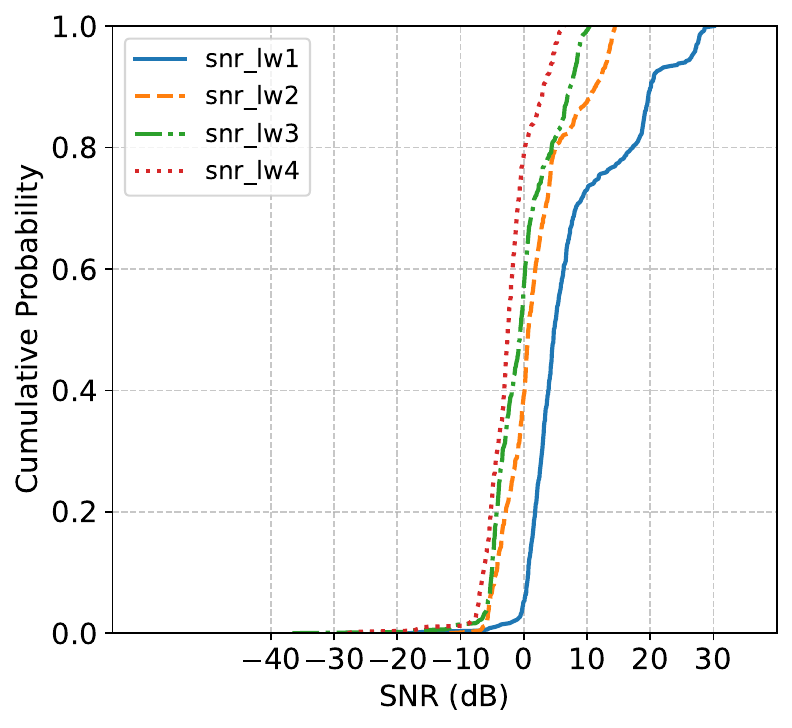}} 
    \hspace{0.2in}
    \subfloat[]{\includegraphics[width=2.8in]{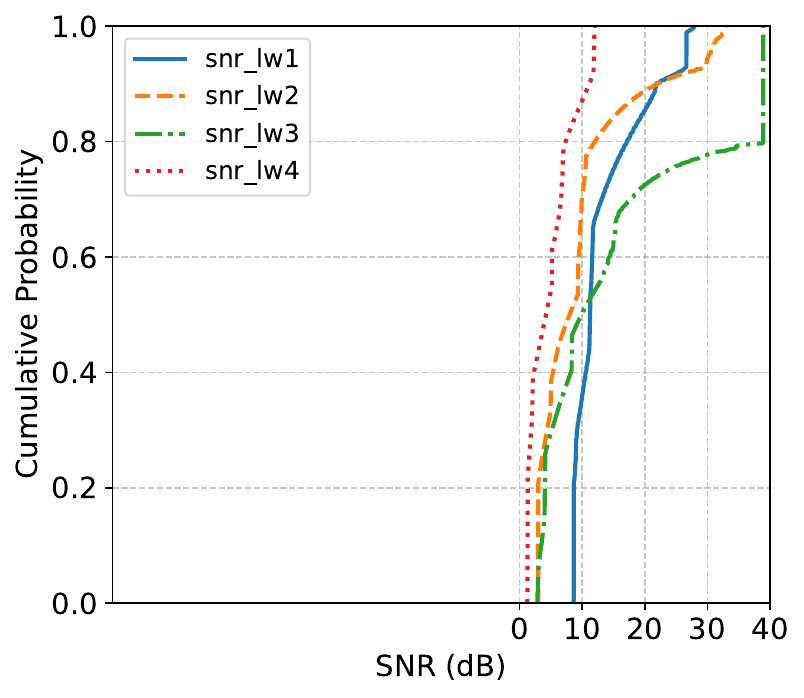}} 
    

    \subfloat[]{\includegraphics[width=2.7in]{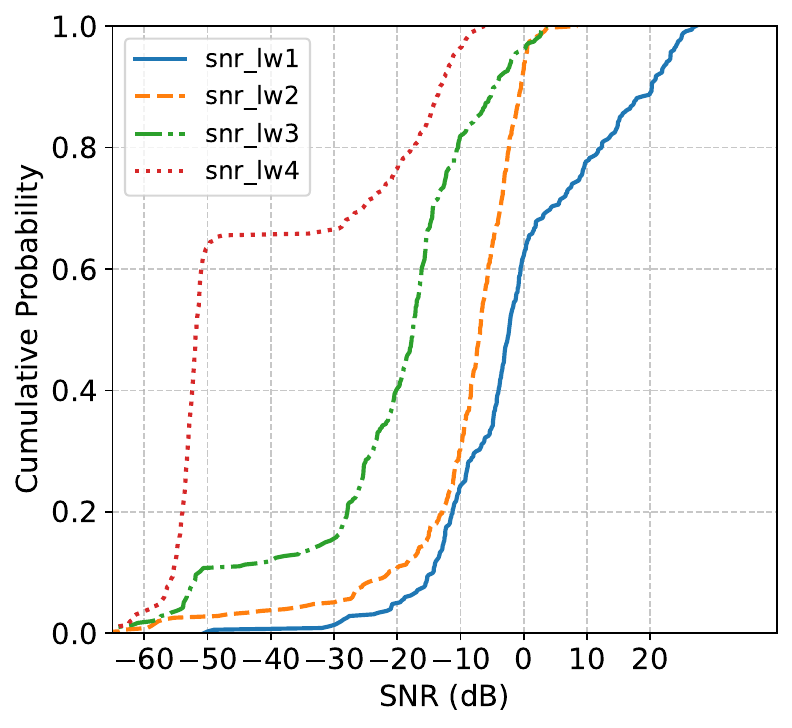}} 
    \hspace{0.2in}
    \subfloat[]{\includegraphics[width=2.7in]{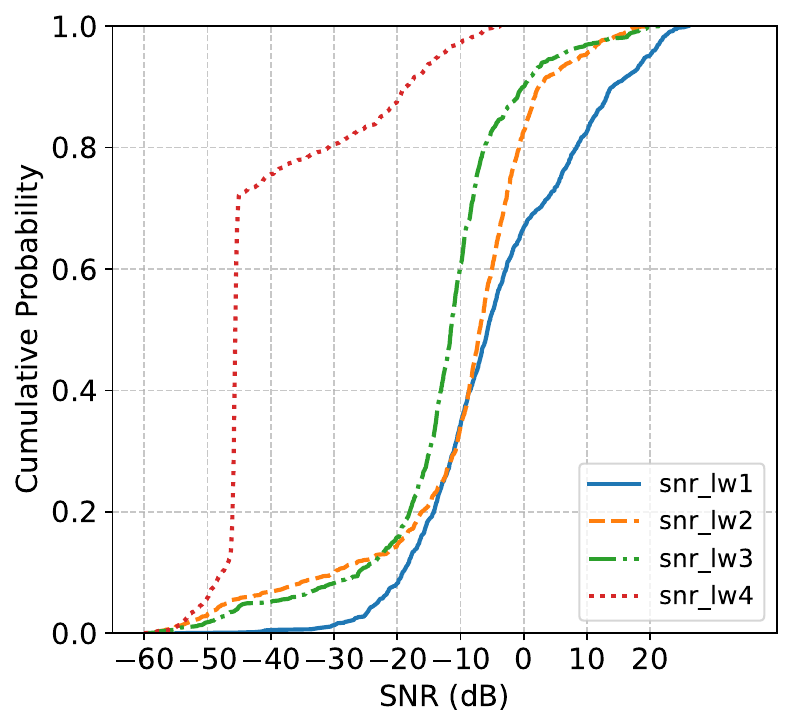}} 
    
    \caption{Distribution of SNR values across four BSs based on (a) fixed trajectory in DT (Fig.~\ref{fig:trajectory}(a)), (b) autonomous trajectory in simulation (Fig.~\ref{fig:trajectory}(b)), (c) AERPAW field testbed-Flight~1, and (d) AERPAW field testbed-Flight~2.}
    \label{fig:cdf_snr_DT_Sim_testbed}
\end{figure*}

Through the experiments, we collect the simulated, DT, and AERPAW RW testbed data to evaluate \alg{HGAD} and baseline \alg{Greedy} scheduling strategies. The experiments capture sensor locations and wireless conditions from the AERPAW testbed~\cite{aerpaw}. The dataset includes time-stamped UAV GPS positions, SNR readings from four sensors, and derived throughput using a standard modulation and coding scheme (MCS) table. For DT scenarios, the distribution of SNR values across four BSs (LW1, LW2, LW3, LW4) is shown in  Fig.~\ref{fig:cdf_snr_DT_Sim_testbed}(a). LW1 consistently provides the best signal quality, as most of its SNR values exceed 10 dB, suggesting a strong and consistent link. LW4 has the poorest SNR values. We see LW2 and LW3 with mild signal fluctuation and varying SNR ranges. Strong transitions in the cumulative distribution function (CDF) curves, especially for LW1 and LW2, indicate that signals behave steadily over time in the real world.

\begin{figure*}[t]
    \centering
    \subfloat[]{\includegraphics[width=3.2in]{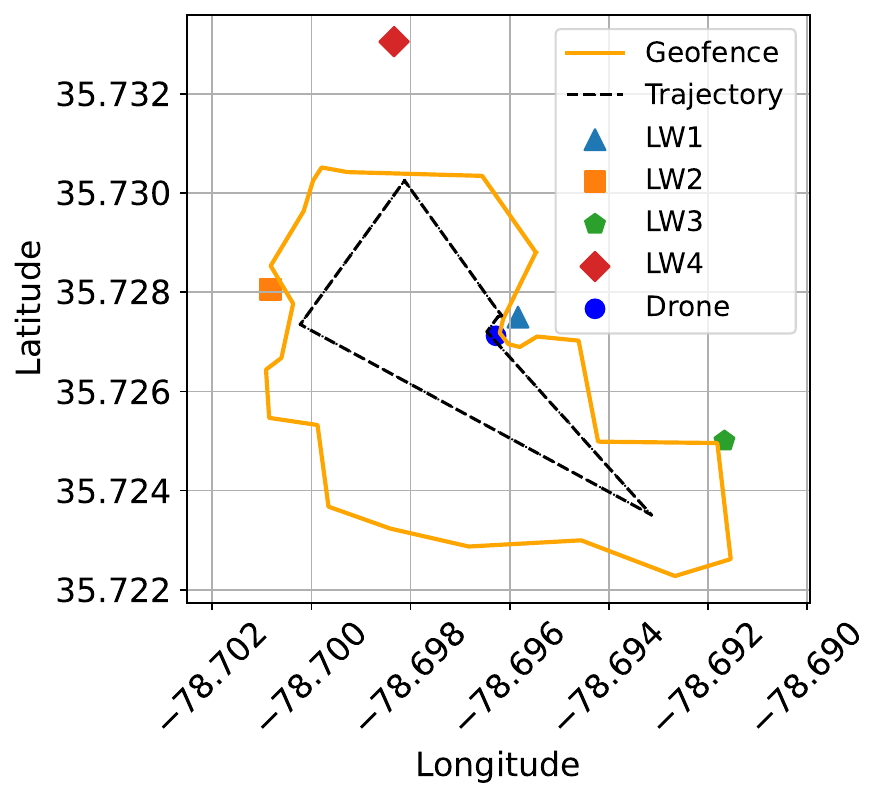}} 
    \subfloat[]{\includegraphics[width=3.2in]{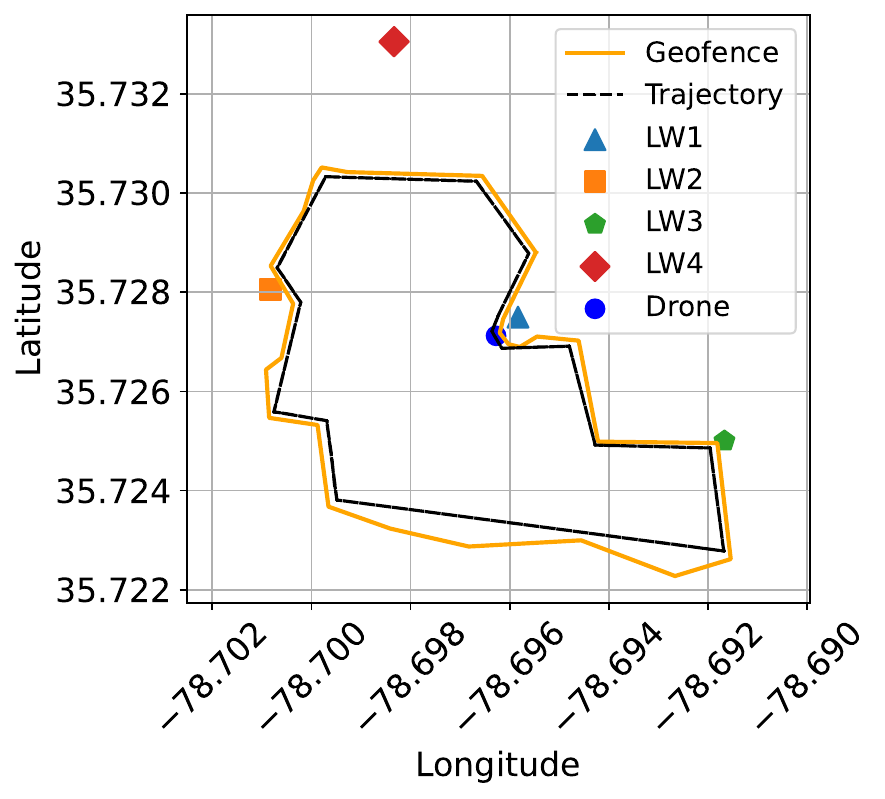}} 
    \caption{UAV fixed trajectory flight missions in AERPAW testbed: (a) Flight 1 and (b) Flight 2.}
    \label{fig:trajectory_testbed}
\end{figure*}

However, the SNR distribution derived from a simulated autonomous trajectory is displayed in Fig.~\ref{fig:cdf_snr_DT_Sim_testbed}(b). Overall, with stronger SNR levels and less severe attenuation, all four BSs appear to offer more optimistic SNR profiles in this instance compared to the AERPAW DT (Fig.~\ref{fig:cdf_snr_DT_Sim_testbed}(a)). A possible discrepancy between modeled and actual conditions is that LW4 performs noticeably better in the simulated environment than in the measured data. Additionally, a broader spread and greater fluctuation in signal strength across the simulated path are shown by the CDF curves' more gradual slopes, especially for LW3 and LW4. To prevent overestimating performance under idealized simulation, this comparison emphasizes the importance of basing UAV scheduling algorithms on the realistic DT traces.

To validate the SNR distributions under realistic wireless scenarios, we undertook two UAV flight campaigns within the AERPAW testbed shown in Fig.~\ref{fig:trajectory_testbed}. They employed pre-specified paths within the geofence. Flight~1 (Fig.~\ref{fig:trajectory_testbed}(a)) is a dense trajectory closer to LW1 and LW2, while Flight~2 (Fig.~\ref{fig:trajectory_testbed}(b)) employs a larger areal coverage with longer proximity closer to all the BSs. This impact is reflected in their corresponding distributions of cumulative SNR shown in Fig.~\ref{fig:cdf_snr_DT_Sim_testbed}(c) and Fig.~\ref{fig:cdf_snr_DT_Sim_testbed}(d). In Flight~1, LW1 is always the strongest contributor with the majority of its SNR values greater than 0~dB (Fig.~\ref{fig:cdf_snr_DT_Sim_testbed}(c)). LW2 and LW3 contribute moderately with greater variability, and LW4 is always the weakest, with almost all values below 0~dB. Flight~2 changes the relative distributions of the links because UAV reaches very close to the BSs, and LW4 is still weak but shows slightly better performance than Flight 1 (Fig.~\ref{fig:cdf_snr_DT_Sim_testbed}(d)).

Comparing DT and simulated outcomes (Fig.~\ref{fig:cdf_snr_DT_Sim_testbed}(a) and Fig.~\ref{fig:cdf_snr_DT_Sim_testbed}(b), respectively), several essential gaps are observed. Simulation results hold the most favorable profiles and overemphasize SNR levels at all BSs, and particularly LW4 performance. DT results strike a balance and correctly discern BS ranking (LW1 strongest and LW4 lowest) and realistic approximations but fail to replicate complete |RW trace variability. RW testbed flights show a significant influence on the empirical distribution of link SNR (Fig.~\ref{fig:cdf_snr_DT_Sim_testbed}(c) and Fig.~\ref{fig:cdf_snr_DT_Sim_testbed}(d)) compared to DT and simulation. 
These findings validate the need to calibrate UAV scheduling strategies, such as \alg{HGAD} under RW testbed scenarios, to account for link variability, environmental uncertainties, and trajectory-dependent variations.

\section{Results and Discussion}\label{sec:nr}
In this section, we examine the performance of \alg{Greedy} and \alg{HGAD} download approaches in four cases: (i) fixed path with DT signal traces, (ii) fixed path with simulation, (iii) an autonomous path in simulation, and (iv) fixed path with two different flights in RW. As shown in Table~\ref{tab:data_buffers}, for the DT, simulation, and RW testbed (Flight 1 shown in Fig.~\ref{fig:trajectory_testbed}(a)) experiment setup, we consider that each BS (sensor) has a fixed amount of data in its buffer: LW1: $500$ Mbits, LW2: $800$ Mbits, LW3: $700$ Mbits, and LW4: $1000$ Mbits. We set the mission time as $500$ seconds for both the DT and simulation settings, while for RW testbed Flight~1 (Fig.~\ref{fig:trajectory_testbed}(a)), it is considered $360$ seconds. Within $500$ seconds, the UAV will download the data from each sensor using either the \alg{Greedy} or \alg{HGAD} approaches. On the other hand, in the case of the AERPAW RW testbed for Flight~2, shown in Fig.~\ref{fig:trajectory_testbed}(b) with mission time $1100$ seconds, we set the data volume for LW1: $1500$ Mbits, LW2: $1300$ Mbits, LW3: $1100$ Mbits, and LW4: $200$ Mbits. These buffer sizes act as completion goals and directly influence the UAV's download choices from a sensor under the \alg{Greedy} and \alg{HGAD} approaches for the aforementioned four scenarios.

\begin{table*}[t]
\centering
\caption{Data buffer configuration assigned to each ground node for DT, simulation, and RW testbed experiments.}
\label{tab:data_buffers}
\renewcommand{\arraystretch}{1.15}
\begin{tabular}{|c|c|c|c|c|c|}
\hline
\textbf{Scenario} & \textbf{Mission time (s)} & \textbf{LW1 (Mbits)} & \textbf{LW2 (Mbits)} & \textbf{LW3 (Mbits)} & \textbf{LW4 (Mbits)} \\
\hline
DT & 500 & 500 & 800 & 700 & 1000 \\
\hline
Simulation & 500 & 500 & 800 & 700 & 1000 \\
\hline
RW Flight 1 & 360  & 500 & 800 & 700 & 1000 \\
\hline
RW Flight 2 & 1100 & 1500 & 1300 & 1100 & 200 \\
\hline
\end{tabular}
\end{table*}

\subsection{Data Mule Operation under Fixed Trajectory with DT Signal Traces}
\begin{figure*}[t]
    \centering
    \subfloat[]{\includegraphics[width=3.3in]{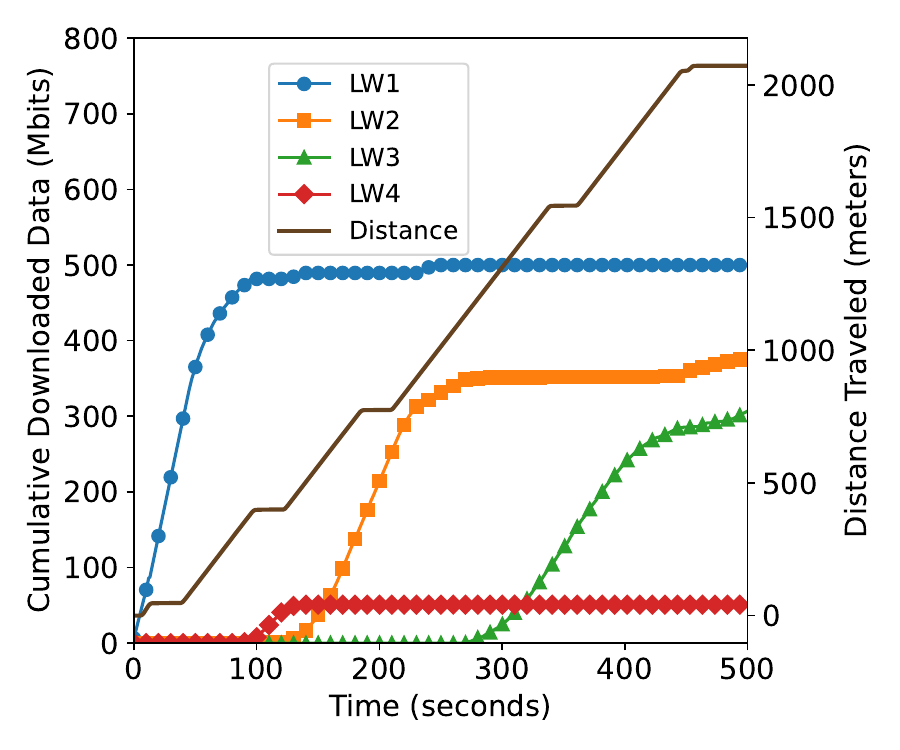}} 
    \hspace{0.2in}
    \subfloat[]{\includegraphics[width=3.3in]{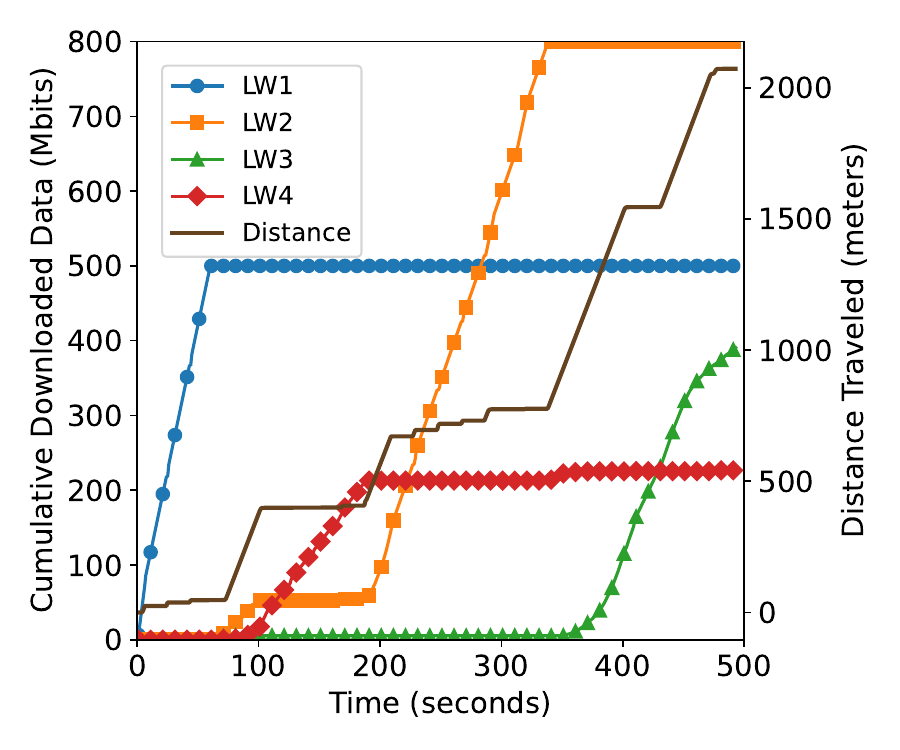}} 
    
    \subfloat[]{\includegraphics[width=3.3in]{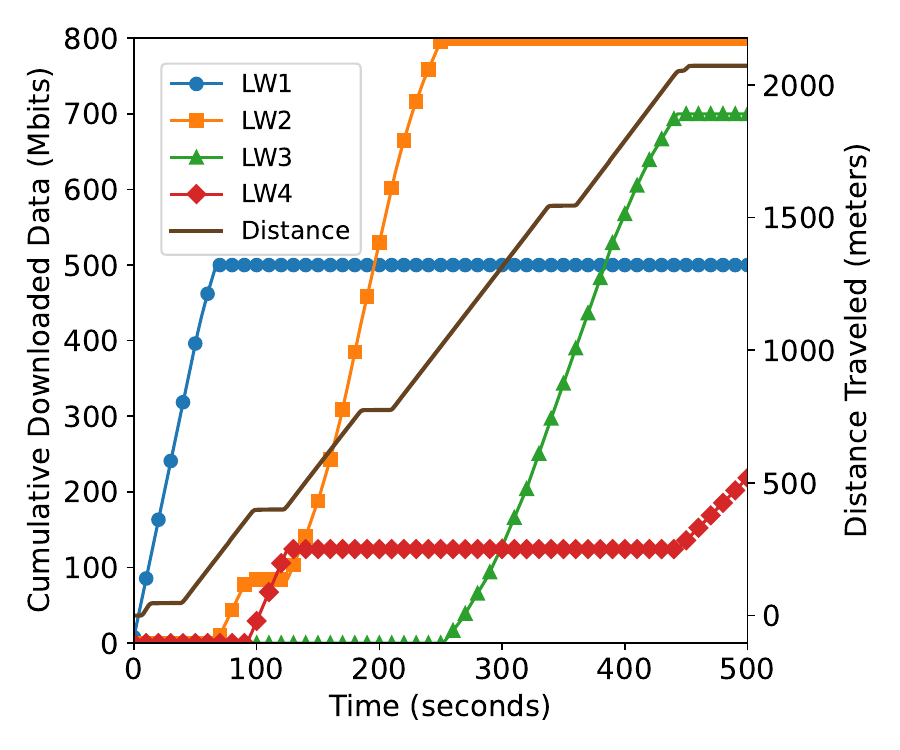}} 
    \hspace{0.2in}
    \subfloat[]{\includegraphics[width=3.3in]{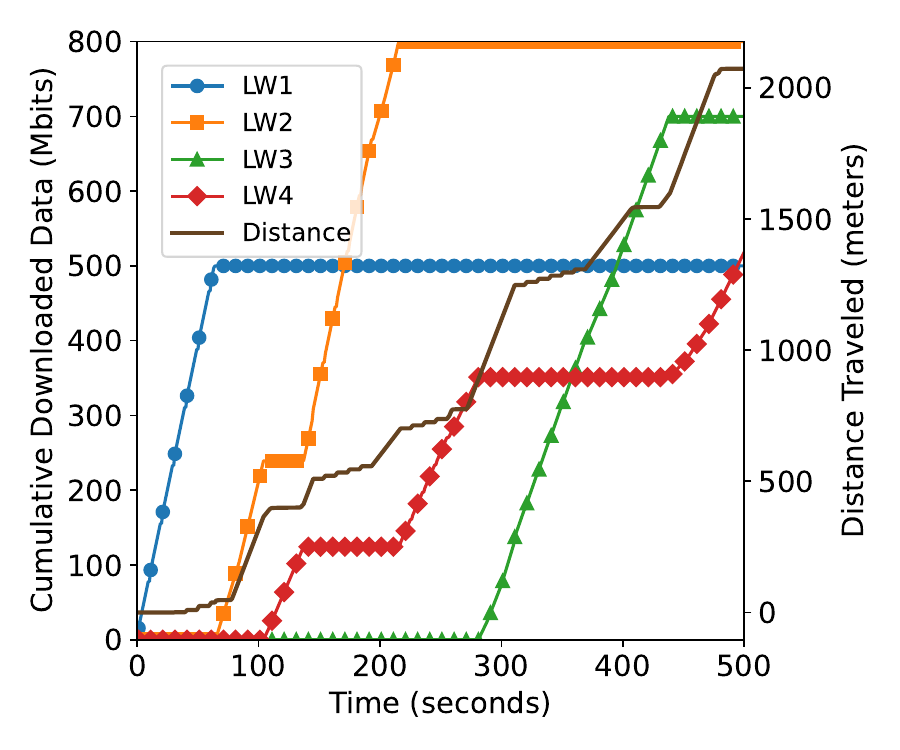}}
    
    \caption{Cumulative downloaded data from each BS over time using (a) \alg{Greedy} approach for DT, (b) \alg{HGAD} for DT, (c) \alg{Greedy} approach for simulation, and (d) \alg{HGAD} for simulation based on a fixed trajectory scenario.}    
    \label{fig:cum_down_fix_DT_sim}
\end{figure*}

Fig.~\ref{fig:cum_down_fix_DT_sim}(a) and Fig.~\ref{fig:cum_down_fix_DT_sim}(b) show the total amount of downloaded data and the UAV distance traveled over time for the DT scenario with two sensor or BS selection approaches. In a \alg{Greedy} strategy shown in Fig.~\ref{fig:cum_down_fix_DT_sim}(a), the UAV frequently switches between BSs, always selecting the one with the highest instantaneous SNR. This results in several transitions, particularly between LWs $1–3$. The UAV mostly downloads from LW1 at the beginning of the mission, then LW2 and LW3, with only a short link to LW4. The brown line (which depicts traveled distance) rises rapidly and frequently, showing that the UAV is always moving and inefficiently downloading data from every BS. This constant movement not only prolongs the mission but also consumes more energy because the UAV spends more time switching between short-lived connections than remaining static under optimal conditions.

In contrast, the \alg{HGAD} approach shown in Fig.~\ref{fig:cum_down_fix_DT_sim}(b) shows a more stable and efficient pattern. The UAV connects to LW1 first and then hovers there for a short period, downloading data at maximum throughput without switching to another sensor to download data. Then, the UAV downloads from other sensors, giving priority to the strongest connection at each stage. It leads to a longer hovering time, fewer sensor transitions, and higher data download.

\subsection{Data Mule Operation under Fixed Trajectory with Simulated Signal Traces}

For the simulated fixed trajectory scenario, Fig.~\ref{fig:cum_down_fix_DT_sim}(c) and Fig.~\ref{fig:cum_down_fix_DT_sim}(d) show the total data download and the distance flown. With early saturation at LW1, the \alg{Greedy} approach aggressively links to the BS with the highest instantaneous SNR in Fig.~\ref{fig:cum_down_fix_DT_sim}(c), generating frequent switching between LW2, LW3, and LW4. This leads to inefficient data download and unnecessary UAV movement. By comparison, Fig.~\ref{fig:cum_down_fix_DT_sim}(d) indicates that \alg{HGAD} downloads in a more ordered manner. After completing downloads from high-SNR BSs, the UAV prioritizes LW4 and LW2 depending on both SNR and remaining data quotas. This leads to increased hover times at specific BSs, smoother cumulative download curves, and a more gradual increase in distance traveled. This results in reduced redundant motion and improved buffer satisfaction by \alg{HGAD}.

\subsection{Data Mule Operation under Autonomous Trajectory (Simulation-Only)}
\begin{figure*}[!t]
    \centering
    \subfloat[BS Selection: LW4, LW3, LW1, LW2]{\includegraphics[width=0.5\linewidth]{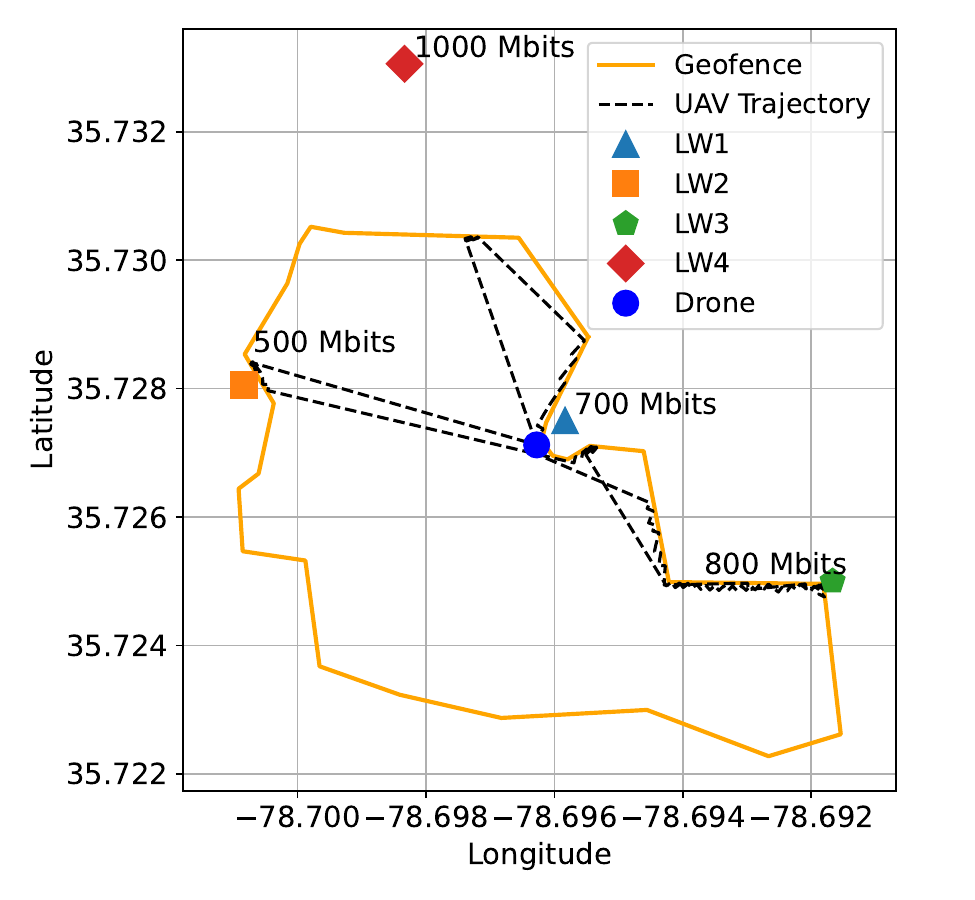}}
    \subfloat[BS Selection: LW3, LW2, LW4, LW1]{\includegraphics[width=0.5\linewidth]{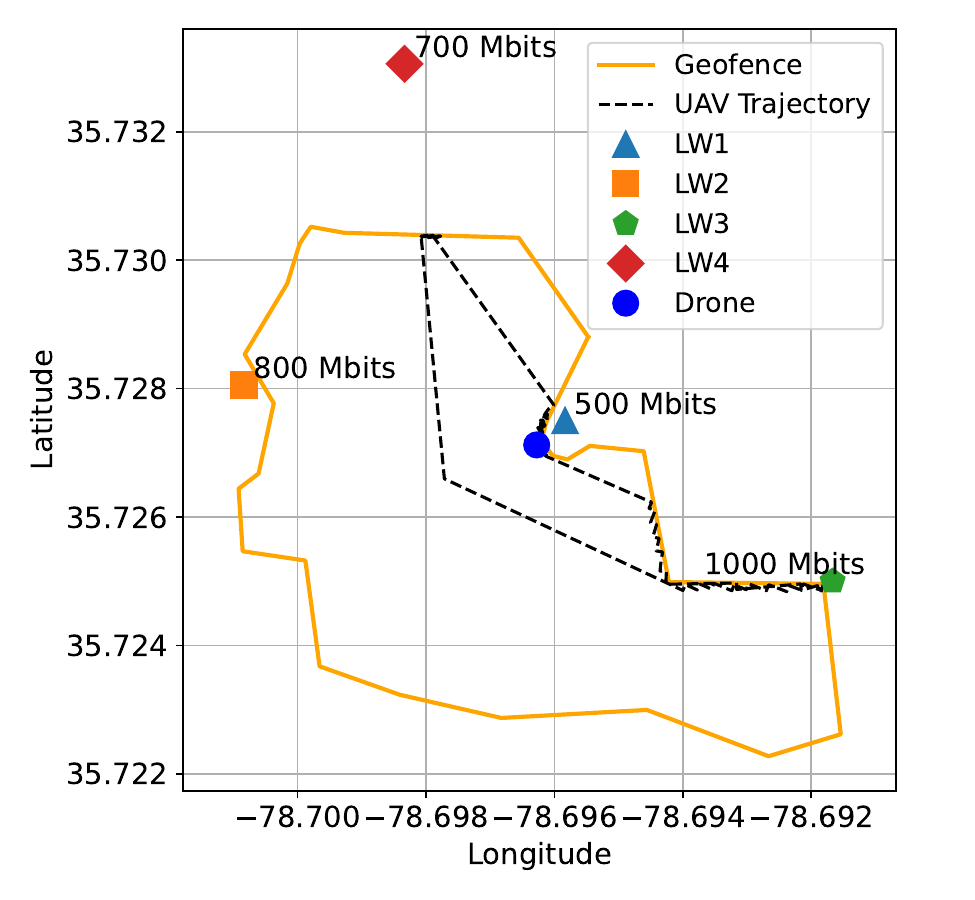}} \\
    \subfloat[BS Selection: LW4, LW1, LW3, LW2]{\includegraphics[width=0.5\linewidth]{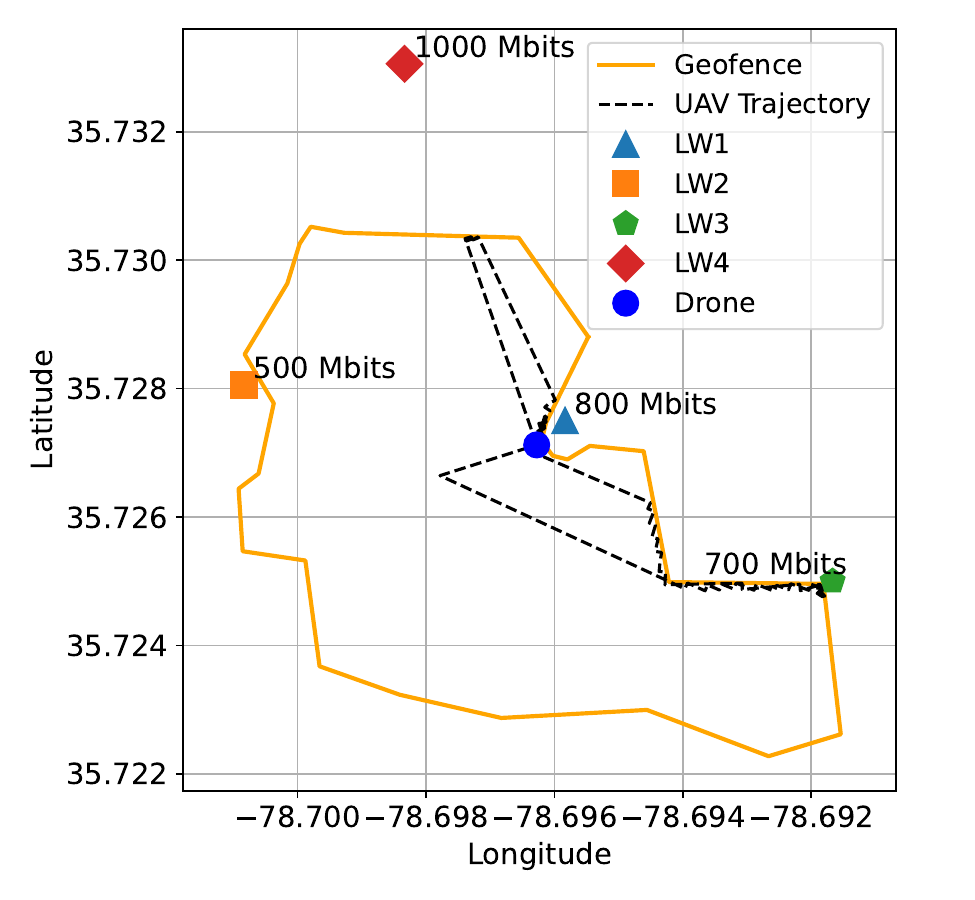}}
    \subfloat[BS Selection: LW2, LW4, LW3, LW1]{\includegraphics[width=0.5\linewidth]{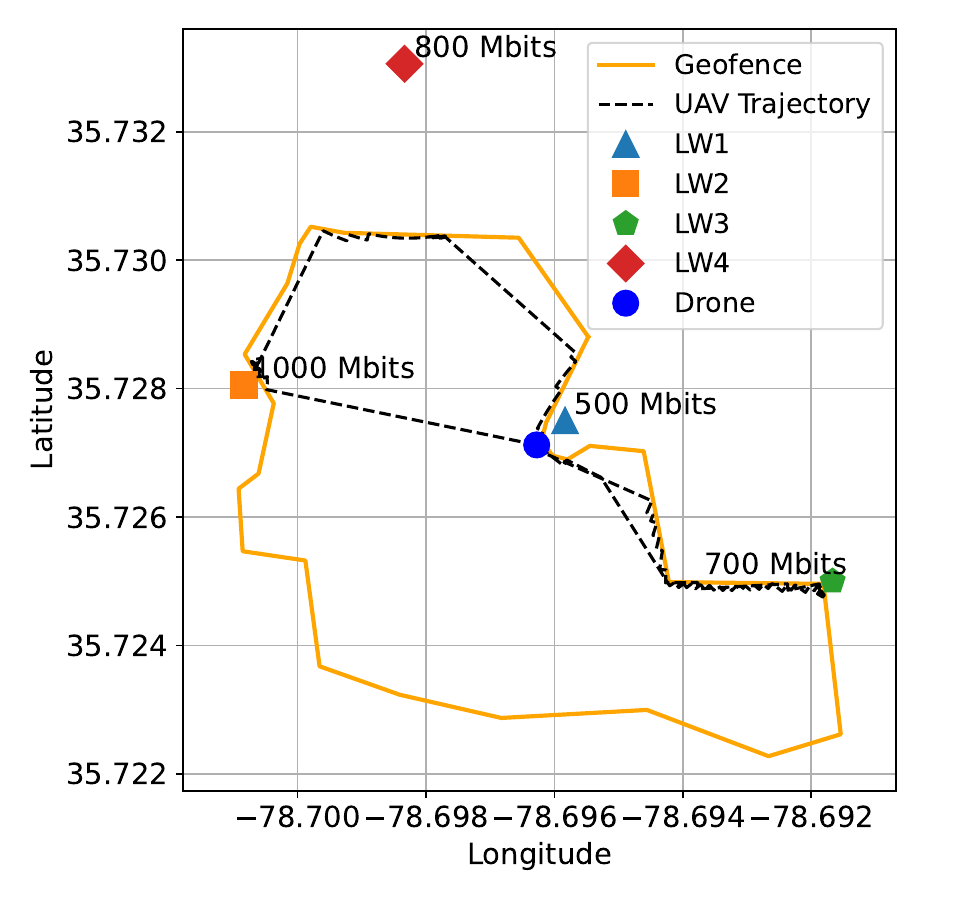}}
    \caption{Examples of four autonomous UAV trajectories.}
    \label{fig:aut_traj_examples}
\end{figure*}

In the simulation environment, we present examples of four autonomous UAV trajectories, as shown in Fig.~\ref{fig:aut_traj_examples}, each illustrating how the UAV adaptively chooses BSs based on the larger data buffers. Table~\ref{tab:fig7_buffers_orders} summarizes the buffer assignments used in Fig.~\ref{fig:aut_traj_examples} and the corresponding BS visitation orders. Each subfigure uses a different buffer configuration, which yields different flight paths. The UAV first moves toward the BS with the largest buffer (e.g., LW4 in Fig.~\ref{fig:aut_traj_examples}(a) and Fig.~\ref{fig:aut_traj_examples}(c), LW3 in Fig.~\ref{fig:aut_traj_examples}(b), and LW2 in Fig.~\ref{fig:aut_traj_examples}(d)), and then sequentially navigates to the next-highest buffer BS. 


\begin{table*}[t]
\centering
\caption{Autonomous UAV trajectories in Fig.~7: data buffers (Mbits) and resulting BS visitation order.}
\label{tab:fig7_buffers_orders}
\renewcommand{\arraystretch}{1.15}
\begin{tabular}{|c|c|c|c|}
\hline
\textbf{Fig.~7} & \textbf{Data buffers (Mbits)} & \textbf{Priority by buffer} & \textbf{UAV navigation order} \\
\hline
(a) & LW1 = 700, LW2 = 500, LW3 = 800, LW4 = 1000 & LW4 $>$ LW3 $>$ LW1 $>$ LW2 & 
LW4$\rightarrow$LW3$\rightarrow$LW1$\rightarrow$LW2 \\
\hline
(b) & LW1 = 500, LW2 = 800, LW3 = 1000, LW4 = 700 & LW3 $>$ LW2 $>$ LW4 $>$ LW1 & LW3$\rightarrow$LW2$\rightarrow$LW4$\rightarrow$LW1 \\
\hline
(c) & LW1 = 800, LW2 = 500, LW3 = 700, LW4 = 1000 & LW4 $>$ LW1 $>$ LW3 $>$ LW2 & LW4$\rightarrow$LW1$\rightarrow$LW3$\rightarrow$LW2 \\
\hline
(d) & LW1 = 500, LW2 = 1000, LW3 = 700, LW4 = 800 & LW2 $>$ LW4 $>$ LW3 $>$ LW1 & LW2$\rightarrow$LW4$\rightarrow$LW3$\rightarrow$LW1 \\
\hline
\end{tabular}
\end{table*}

\begin{figure*}[t]
    \centering
    \subfloat[]{\includegraphics[width=3.2in]{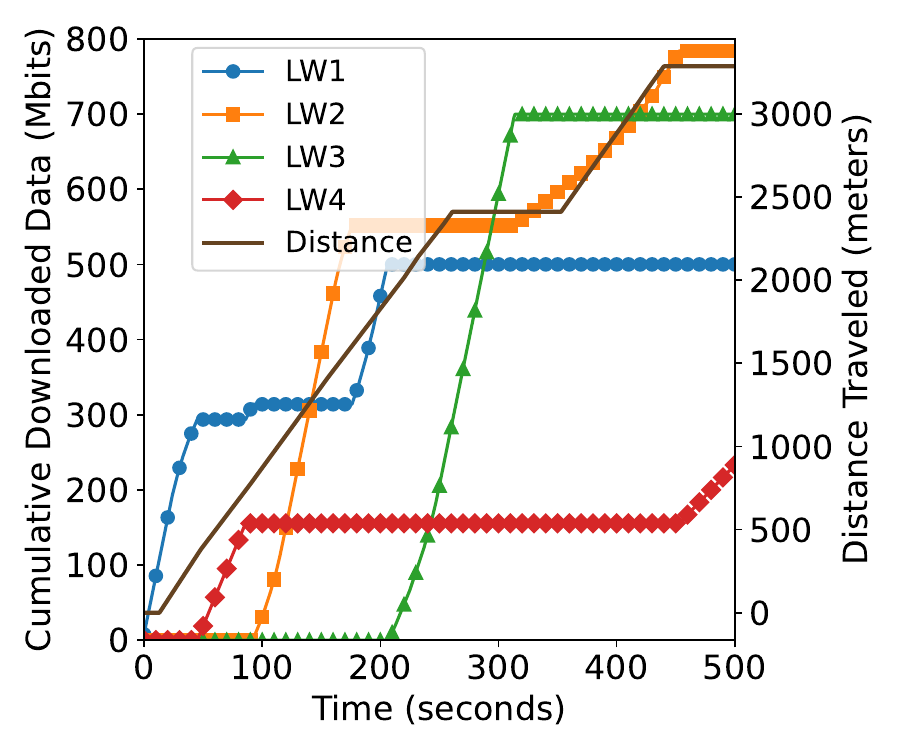}} 
    \hspace{0.2in}
    \subfloat[]{\includegraphics[width=3.2in]{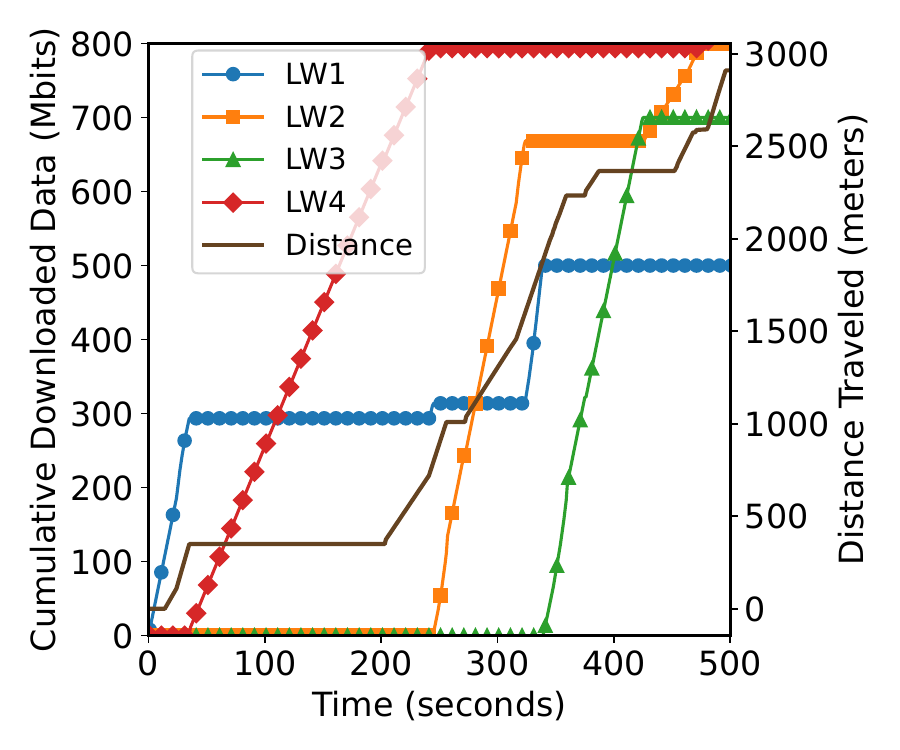}} 
    \hspace{0.2in}
    \subfloat[]{\includegraphics[width=3.2in]{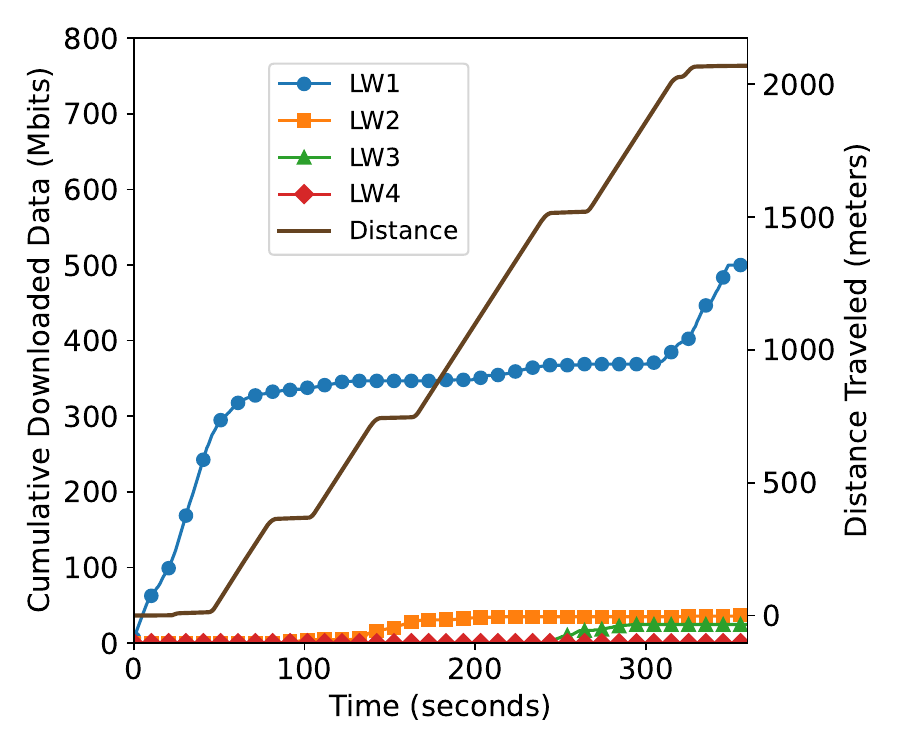}} 
    \hspace{0.2in}
    \vspace{-.1in}
    \subfloat[]{\includegraphics[width=3.2in]{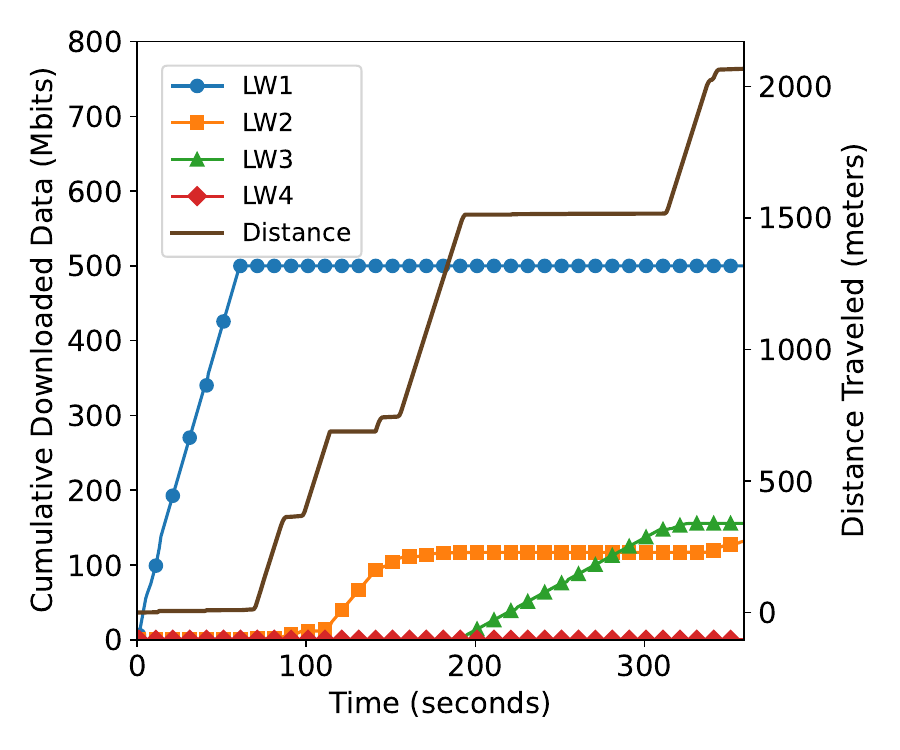}} 

    \subfloat[]{\includegraphics[width=3.2in]{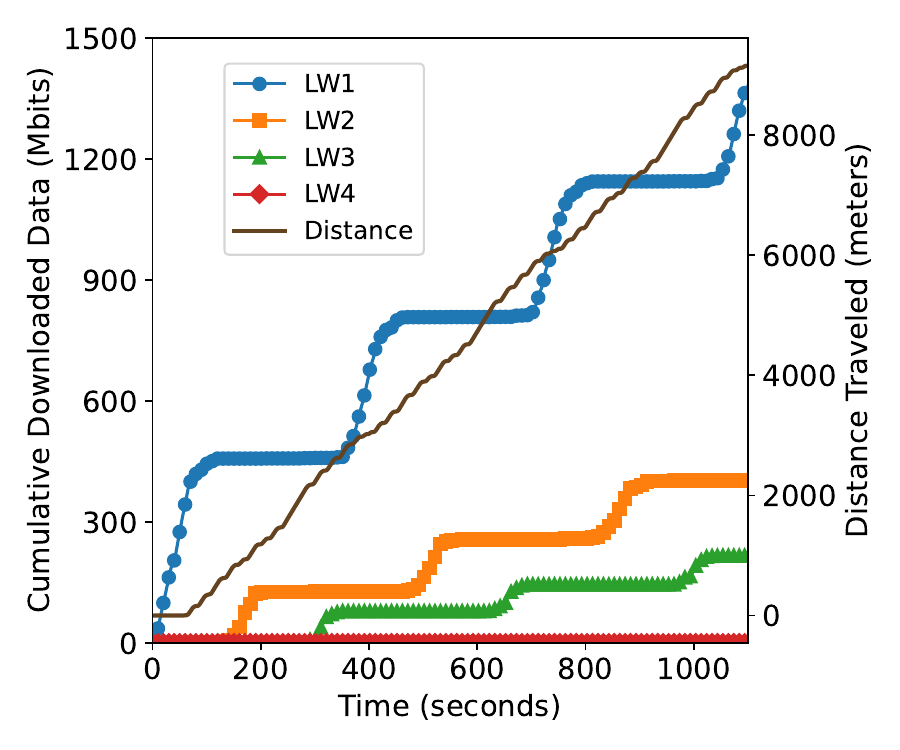}} 
    \hspace{0.2in}
    \subfloat[]{\includegraphics[width=3.2in]{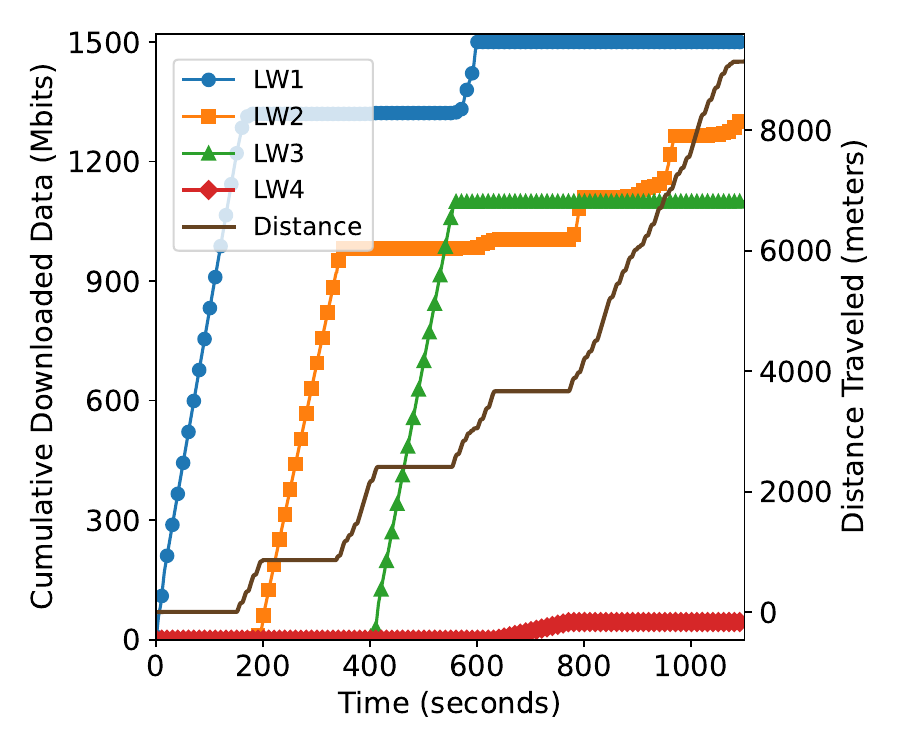}} 
    \caption{Cumulative downloaded data from each BS over time using (a) \alg{Greedy} and (b) \alg{HGAD} approaches for the simulated autonomous trajectory scenario; (c) \alg{Greedy} and (d) \alg{HGAD} approaches for the AERPAW field testbed for Flight~1; (e) \alg{Greedy} and (f) \alg{HGAD} approaches for the AERPAW field testbed for Flight~2.}
    \label{fig:cum_down_testbed_sim-aut_field-flight1-2}
\end{figure*}

Although we show some examples (Fig.~\ref{fig:aut_traj_examples}) to validate the adaptability of the autonomous trajectory algorithm, we consider the autonomous trajectory mentioned in Fig.~\ref{fig:trajectory}(b) for the simulation experiment. In contrast to the \alg{Greedy} approach shown in Fig.~\ref{fig:cum_down_testbed_sim-aut_field-flight1-2}(a), which significantly favors LW1 and demonstrates early saturation, the \alg{HGAD} strategy illustrated in Fig.~\ref {fig:cum_down_testbed_sim-aut_field-flight1-2}(b) allows more efficient and balanced buffer fulfillment across all BSs, particularly LW4. Although the UAV travels a similar total distance, \alg{HGAD} achieves a higher total data download by employing more effective path planning, resulting in smoother download transitions.

\subsection{Data Mule Operation under Fixed  Trajectory with AERPAW Real-World Testbed}

In terms of the RW testbed experiment, we compare the total data download between \alg{Greedy} and \alg{HGAD} strategies, shown in Fig.~\ref{fig:cum_down_testbed_sim-aut_field-flight1-2}(c)-Fig.~\ref{fig:cum_down_testbed_sim-aut_field-flight1-2}(d). For both RW AERPAW flights, \alg{HGAD} consistently outperforms the \alg{Greedy} baseline for downloading more data from the BSs. From Table~\ref{tab:download_comparison}, for Flight~1 (Fig.~\ref{fig:trajectory_testbed}(a)), we considered a fixed dense trajectory closer to LW1 and LW2. Here, the \alg{Greedy} approach (Fig.~\ref{fig:cum_down_testbed_sim-aut_field-flight1-2}(c)) resulted in only about $563$ Mbits of data download, whereas \alg{HGAD} (Fig.~\ref{fig:cum_down_testbed_sim-aut_field-flight1-2}(d)) resulted in nearly $787$ Mbits of data download with an improvement of around $40$\%.~For Flight~2 (Fig.~\ref{fig:trajectory_testbed}(b), the UAV trajectory covered a bigger area with longer proximity to all BSs, where \alg{Greedy} approach (Fig.~\ref{fig:cum_down_testbed_sim-aut_field-flight1-2}(e)) showed downloading $2002$ Mbits of data, while \alg{HGAD} (Fig.~\ref{fig:cum_down_testbed_sim-aut_field-flight1-2}(f)) doubled this with $3944$ Mbits, showing approximately a $97$\% improvement. These results indicate that \alg{HGAD}'s hover-and-buffer-aware approach enables more balanced data collection across BSs even under RW wireless channel conditions, while \alg{Greedy} tends to overcommit to LW1 and neglects weaker but quota-constrained BSs.

Overall, as illustrated in Table~\ref{tab:download_comparison}, the \alg{HGAD} outperforms the \alg{Greedy} approach in terms of the total data download. First, the simulation environment consistently shows greater total data download for both strategies due to idealized channel models and also overstates LW4's effectiveness. Here, \alg{HGAD} outperforms \alg{Greedy} by $14$\% on fixed trajectories and $29$\% on autonomous trajectories. Second, although the DT shows the signal strength between simulation and RW, it accurately reflects the BS ranking, i.e., LW1 is the strongest and LW4 the weakest. It shows lower total data download than the simulation but more realistic patterns. Here, \alg{HGAD} outperforms \alg{Greedy} by $57$\% on the fixed trajectory, reflecting better stability under fading circumstances. Finally, as the RW flights include hardware constraints, multipath, and UAV dynamics, these flights capture the real-world environments. Hence, we observe lower throughput than DT and simulation. However, for all cases, the simulation, DT, and RW testbeds, we notice a similar performance of \alg{HGAD} compared to \alg{Greedy} in terms of the total data download. Overall, these results demonstrate that although simulation is useful for prototyping, DT is still a reliable intermediate validation step that maintains realistic link orderings and trends. And RW testbed trials ultimately validate \alg{HGAD}’s resilience and practical importance in RW deployments.

\begin{table*}[h!]
\centering
\caption{Comparison of total data download under different scenarios.}
\label{tab:download_comparison}
\begin{tabular}{|l|c|c|c|c|}
\hline
\textbf{Scenario} & \textbf{Trajectory}& \textbf{Greedy (Mbits)} & \textbf{HGAD (Mbits)} & \textbf{Time (s)} \\
\hline
DT & Fixed & 1233.14 & 1929.78 & 500\\
\hline
Simulation & Fixed  & 2218.5 & 2518.18 & 500\\
\hline
Simulation & Autonomous & 2223.1 & 2863.9 & 500\\
\hline
RW & Fixed (Flight 1) & 562.75 & 787.44 & 360\\
\hline
RW & Fixed (Flight 2) & 2001.52 & 3944.324 & 1100\\
\hline
\end{tabular}
\end{table*}


\section{Conclusion and Future Work}\label{sec:con}
In this paper, we present a resilient framework for UAV-based wireless data collection in mission-restricted areas where the UAV operates along predetermined and autonomous trajectories with tight time and sensor buffer constraints. By using realistic signal traces from the NSF AERPAW DT and RW testbed along with simulation development, we compare two strategies: a traditional \alg{Greedy} heuristic and a \alg{Hover-based Greedy Adaptive Download (HGAD)} strategy for sensor selection. While the \alg{Greedy} algorithm abruptly moves to the sensor with the higher instantaneous signal-to-noise ratio (SNR), \alg{HGAD} provides a mechanism of stability that enables the UAV to stop and resume downloading data when peak throughput conditions are found. This hover-based logic enhances download stability, improves download satisfaction, and minimizes unnecessary movement, resulting in increased mission efficiency and utilization of mission time and energy. Our results demonstrate that \alg{HGAD} provides greater sensor data buffer satisfaction and higher total data download than the \alg{Greedy} heuristic. These advantages make \alg{HGAD} particularly suitable for tactical and emergency operations, where UAV autonomy, efficiency, and reliability are vital. Such strategies can directly support emergency response UAVs tasked with resilient data collection from IoT sensors in disaster zones, where connectivity and mission time are highly constrained.

For future work, we plan to extend this research by proposing a novel framework for optimizing the UAV trajectory using a reinforcement learning algorithm with the integration of a neural network-based digital twin, minimizing the gap between the simulation and real-world data. Additionally, we plan to accommodate multiple UAVs to ease the offloading of information from disparate sensors without overlapping.

\acknowledgements
This work is supported in part by the NSF award CNS-1939334, and in part by the NASA ULI award 80NSSC25M7102.

\bibliographystyle{IEEEtran}   
\bibliography{ref}            
\thebiography
\begin{biographywithpic}
{Md Sharif Hossen}{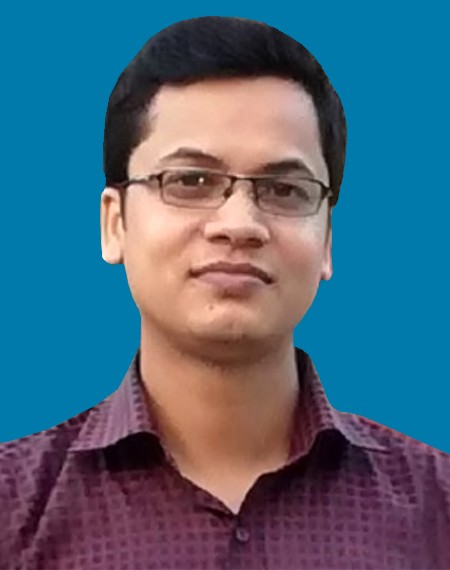} 
(Senior Member, IEEE) received his B.Sc. and M.Sc. degrees in Information and Communication Engineering from the University of Rajshahi, Bangladesh. He is an Assistant Professor at Comilla University, Bangladesh, currently on study leave. He is pursuing his Ph.D. in Electrical and Computer Engineering at North Carolina State University. His research focuses on O-RAN, digital twins, UAV path planning, and next-generation wireless networks. He served as a reviewer for several international journals (viz., Springer, IEEE Access, Wiley) and conferences (IEEE NFV-SDN, IEEE WiSec). He received several scholarships, like an ICT research fellowship (for his M.Sc. thesis), UGC scholarship (for the highest B.Sc. result in the faculty), Merit scholarship for outstanding academic results at the University of Rajshahi, and a graduate merit scholarship from North Carolina State University. He received the Best Paper Award at the IEEE ICISET 2016 and at the Springer ICACIE 2018 conference.
\end{biographywithpic} 
\begin{biographywithpic}
{An\i l G\"{u}rses}{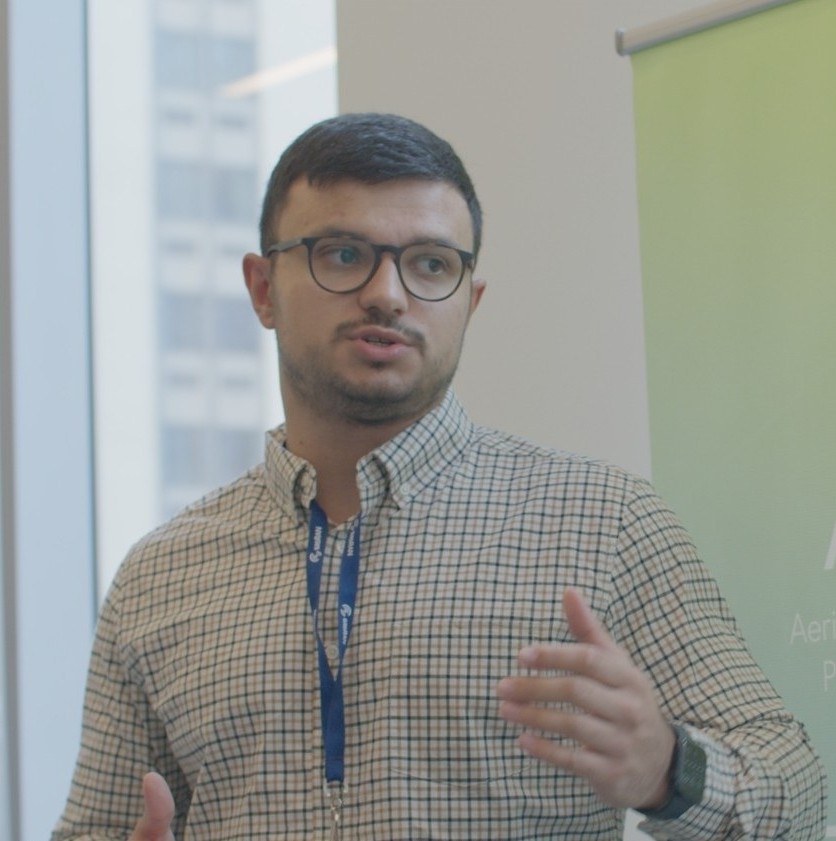}
is a PhD student in Electrical and Computer Engineering at North Carolina State University. He received his B.S. degree in Electrical and Electronics Engineering from Istanbul Medeniyet University, Turkey, in 2021. His research interests include wireless communications, channel modeling, digital twins, and UAV networks.
\end{biographywithpic}
\begin{biographywithpic}
{Ozgur Ozdemir}{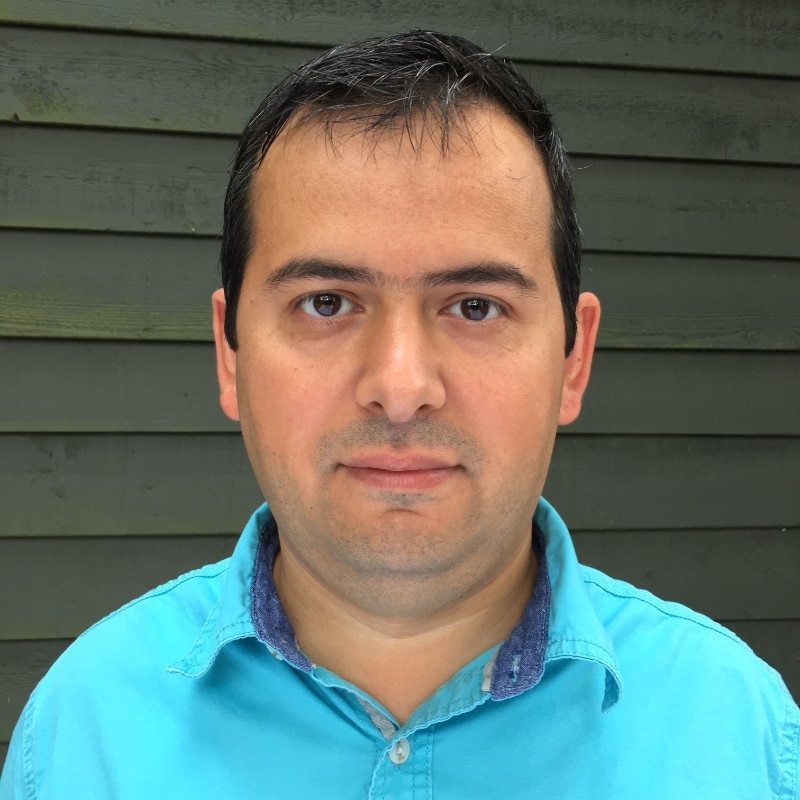} (Member, IEEE) received the B.S. degree in electrical and electronics engineering from Bogazici University, Istanbul, Türkiye, in 1999, and the M.S. and Ph.D. degrees in electrical engineering from The University of Texas at Dallas, Richardson, TX, USA, in 2002 and 2007, respectively. From 2007 to 2016, he was an Assistant Professor at Fatih University, Istanbul, and a Postdoctoral Scholar at Qatar University, Doha, Qatar, for 3.5 years. He joined the Department of Electrical and Computer Engineering, NC State, as a Visiting Research Scholar in 2017. He is currently an Associate Research Professor. He is the lead for supporting field experimentation with wireless technologies and drones for the NSF AERPAW platform at NC State. His research interests include software-defined radios, channel sounding for mmWave systems, wireless testbeds, digital compensation of radio-frequency impairments, and opportunistic approaches in wireless systems.
\end{biographywithpic}

\begin{biographywithpic}
{Mihail L. Sichitiu}{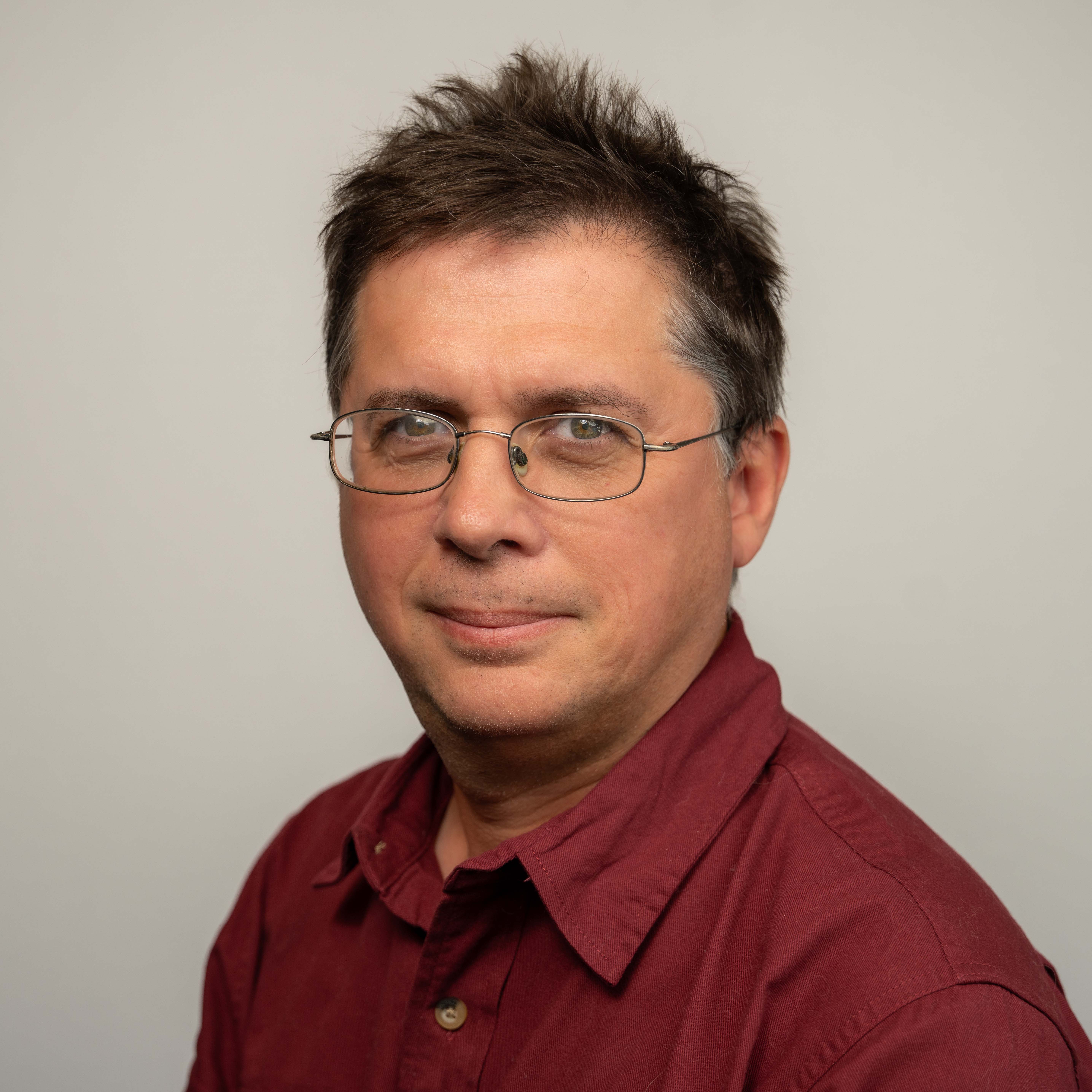} earned his Ph.D. degree in Electrical Engineering from the University of Notre Dame in 2001. His current research interests include wireless networks and communications for UAVs. In these systems, he is studying problems related to localization, time synchronization, emulation, routing, fairness, and modeling. He is teaching wireless networking and UAV courses. He is a professor in the Department of Electrical and Computer Engineering at NCSU.
\end{biographywithpic}

\begin{biographywithpic}
{İsmail Guvenc}{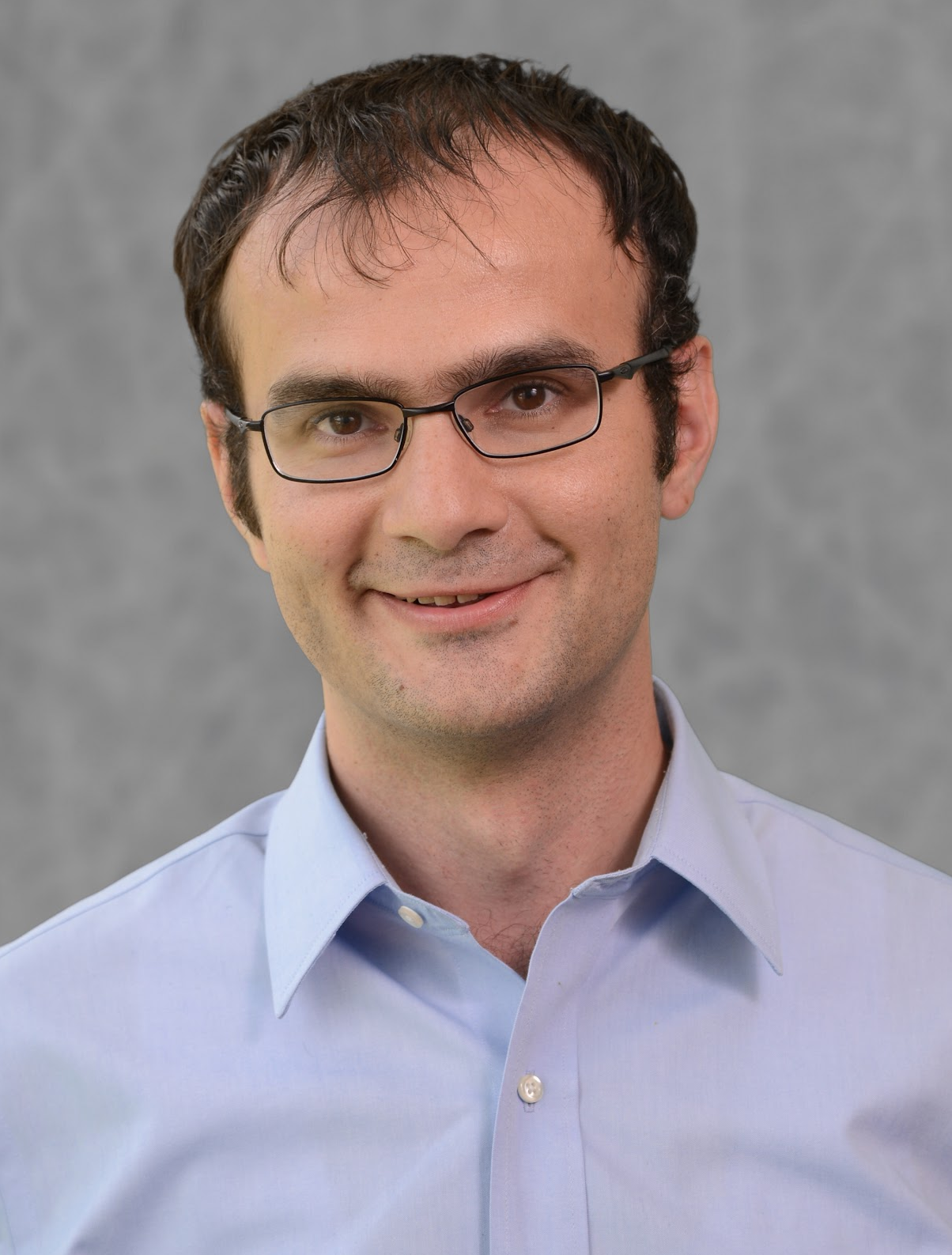} (Fellow, IEEE) received the Ph.D. degree in electrical engineering from the University of South Florida, Tampa, FL, USA, in 2006. He is currently a Professor with the Department of Electrical and Computer Engineering, North Carolina State University, Raleigh, NC, USA. He has authored or coauthored more than 300 conference/journal papers and book chapters, several standardization contributions, four books, and more than 30 U.S. patents. He is the PI and the Director of the NSF AERPAW Project and the Site Director of the NSF BWAC I/UCRC Center. His recent research interests include 5G or 6G wireless networks, UAV communications, millimeter or terahertz communications, and heterogeneous networks.

Prof. Güvenç was the recipient of several awards, including the NC State Faculty Scholar Award in 2021, the R. Ray Bennett Faculty Fellow Award in 2019, the FIU COE Faculty Research Award in 2016, the NSF CAREER Award in 2015, the Ralph E. Powe Junior Faculty Award in 2014, and the USF Outstanding Dissertation Award in 2006. He is a Senior Member of the National Academy of Inventors.
\end{biographywithpic}

\end{document}